\documentclass[final,3p,12pt,authoryear]{elsarticle}

\usepackage{amsmath,amssymb,color,graphicx,sectsty,caption,subcaption,url,ulem,natbib}
\newcommand{\be}{\begin{equation}}
\newcommand{\bea}{\begin{eqnarray}}
\newcommand{\ee}{\end{equation}}
\newcommand{\eea}{\end{eqnarray}}
\newcommand{\trans}{\scriptscriptstyle\mskip-1mu\top\mskip-2mu}
\newcommand{\del}{\delta}
\newcommand{\gogo}{\notag\\[4pt]}

\newcommand{\calC}{\mathcal{C}}

\newcommand{\calS}{\mathcal{S}}
\newcommand{\calP}{\mathcal{P}}

\newcommand{\bfchi}{\boldsymbol{\chi}}
\newcommand{\eps}{\varepsilon}

\newcommand{\grads}{\nabla_{\mskip-2mu\scriptscriptstyle\calS}}

\newcommand{\divs}{\text{div}_{\mskip-2mu\scriptscriptstyle\calS\mskip2mu}}
\newcommand{\tr}{\text{tr}\mskip2mu}

\newcommand{\dr}{\,\text{d}r}

\newcommand{\dtheta}{\,\text{d}\theta}

\newcommand{\drdtheta}{\,\text{d}r\mskip1mu\text{d}\theta}

\newcommand{\lint}{\int_0^{2\pi}}

\newcommand{\aintdim}{\int_0^{2\pi}\mskip-8mu\int_0^1}
\newcommand{\crig}{c\mskip1.5mu}

\newcommand{\go}{\mskip1.0mu}

\newcommand{\half}{{\textstyle{\frac{1}{2}}}}

\newcommand{\abs}[1]{|#1\rvert}

\newcommand{\bfb}{\boldsymbol{b}}

\newcommand{\bfd}{\boldsymbol{d}}
\newcommand{\bfe}{\boldsymbol{e}}
\newcommand{\bfF}{\boldsymbol{F}}
\newcommand{\bft}{\boldsymbol{t}}
\newcommand{\bfx}{\boldsymbol{x}}
\newcommand{\bfo}{\boldsymbol{o}}

\newcommand{\bfm}{\boldsymbol{m}}
\newcommand{\bfn}{\boldsymbol{n}}

\newcommand{\bfr}{\boldsymbol{r}}
\newcommand{\bfu}{\boldsymbol{u}}

\newcommand{\bfv}{\boldsymbol{v}}
\newcommand{\bfw}{\boldsymbol{w}}

\newcommand{\bfk}{\boldsymbol{\kappa}}
\newcommand{\bfnu}{\boldsymbol{\nu}}

\newcommand{\bfxi}{\boldsymbol{\xi}}

\newcommand{\bfA}{\boldsymbol{A}}
\newcommand{\bfS}{\boldsymbol{S}}

\newcommand{\bsu}{\boldsymbol{\epsilon}}

\newcommand{\xir}{\bfxi_r}

\newcommand{\xit}{\bfxi_{\theta}}
\newcommand{\xitt}{\bfxi_{\theta \theta}}
\newcommand{\xittt}{\bfxi_{\theta \theta \theta}}

\newcommand{\reqone}{\scriptstyle r\mskip1.75mu=\mskip0.75mu1}

\newcommand{\vit}{v_{\theta}}
\newcommand{\vitt}{v_{\theta \theta}}

\newcommand{\wir}{w_{r}}

\newcommand{\wit}{w_{\theta}}
\newcommand{\witt}{w_{\theta \theta}}

\newcommand{\uit}{u_{\theta}}

\newcommand{\ut}{\bfu_{\theta}}
\newcommand{\utt}{\bfu_{\theta \theta}}

\newcommand{\bfP}{\boldsymbol{P}}

\newfont{\tenbss}{bbmss12}

\newcommand{\bsw}{\mbox{\tenbss w}}

\newcommand{\Div}{\hbox{\rm div}\mskip2mu}
\newcommand{\divS}{\Div_{\mskip-3mu\scriptscriptstyle\calS}}

\newfont{\tenbfsl}{cmbxti12 scaled 1050}% <-- for idem tensor
\newcommand{\idem}{\mbox{\tenbfsl 1\/}}

\newcommand{\sperp}{\scriptscriptstyle\perp}
\newcommand{\er}{\bfe}
\newcommand{\et}{\bfe^{\sperp}}
\newcommand{\ez}{\bfe\times\bfe^{\sperp}}
\newcommand{\erthet}{\bfe(\theta)}
\newcommand{\etthet}{\bfe^{\sperp}(\theta)}
\newcommand{\ezthet}{\bfe(\theta)\times\bfe^{\sperp}(\theta)}

\def\clb{\color{blue}}
\def\clr{\color{red}}

\def\citeapos#1{\citeauthor{#1}'s (\citeyear{#1})}

\journal{International Journal of Engineering Science}

\newcommand{\captionfonts}{\footnotesize}

\makeatletter  % Allow the use of @ in command names
\long\def\@makecaption#1#2{%
  \vskip\abovecaptionskip
  \sbox\@tempboxa{{\captionfonts #1: #2}}%
  \ifdim \wd\@tempboxa >\hsize
    {\captionfonts #1: #2\par}
  \else
    \hbox to\hsize{\hfil\box\@tempboxa\hfil}%
  \fi
  \vskip\belowcaptionskip}
\makeatother   % Cancel the effect of \makeatletter

\begin{document}

\begin{frontmatter}

%% Title, authors and addresses

%% use the tnoteref command within \title for footnotes;
%% use the tnotetext command for theassociated footnote;
%% use the fnref command within \author or \address for footnotes;
%% use the fntext command for theassociated footnote;
%% use the corref command within \author for corresponding author footnotes;
%% use the cortext command for theassociated footnote;
%% use the ead command for the email address,
%% and the form \ead[url] for the home page:
%% \title{Title\tnoteref{label1}}
%% \tnotetext[label1]{}
%% \author{Name\corref{cor1}\fnref{label2}}
%% \ead{email address}
%% \ead[url]{home page}
%% \fntext[label2]{}
%% \cortext[cor1]{}
%% \address{Address\fnref{label3}}
%% \fntext[label3]{}

\title{Theoretical and experimental study of the stability of a soap film spanning a flexible loop}

% use optional labels to link authors explicitly to addresses:
 \author[McGill]{Aisa Biria}
 \author[OIST]{Eliot Fried\corref{cor1}}
 \ead{eliot.fried@oist.jp}
  \address[McGill]{Department of Mechanical Engineering, McGill University, Montr\'eal, Qu\'ebec H3A 0C3, Canada }
 \address[OIST]{Mathematical Soft Matter Unit,
Okinawa Institute of Science and Technology Graduate University, Onna, Okinawa 904-0495, Japan}
\cortext[cor1]{Corresponding author}

\begin{abstract}
A variational model is used to study the stability of a soap film spanning a flexible loop. The film is modeled as a fluid surface endowed with constant tension and the loop is modeled as an inextensible and unshearable elastic rod that resists not only bending but also twisting about the tangent to its centerline. The first and second energy variations of the underlying energy functional are derived, leading to governing equilibrium equations and energetically based stability conditions. The latter conditions are applied to explore the effect of subjecting flat circular configurations to planar and transverse perturbations. Both the areal stretching and the lineal twisting prove to compete against the lineal bending to destabilize such a configuration.
Experiments are performed to study post-buckled configurations and to assess the validity of the theoretical predictions. 
\end{abstract}
\begin{keyword}
%% keywords here, in the form: keyword \sep keyword
Stability analysis; soap film; surface tension; minimal surface; vanishing mean curvature; Kirchhoff elastic rod; bending energy; torsional energy
%% PACS codes here, in the form: \PACS code \sep code
%% MSC codes here, in the form: \MSC code \sep code
%% or \MSC[2008] code \sep code (2000 is the default)
\end{keyword}

\end{frontmatter}

%% \linenumbers

%\newpage

\section{Introduction}
\label{introduction}
The thread problem posed by \citet{Alt} concerns minimal surfaces bounded by loops consisting of inextensible segments of prescribed lengths, some of which are rigid and the remainder of which are flexible. The specialization of this problem that arises if the bounding loop contains no rigid segments was first considered by \citet{bernatzki2001}, who modeled the film as a surface with uniform free-energy density and the bounding loop as a curve with bending-energy density quadratic in its curvature. For the complementary limiting case in which the bounding loop is inflexible, the thread problem reduces to the Plateau problem, namely finding the surface of least area that spans a rigid frame of given shape. In the limit of vanishing surface energy, the problem specializes to finding the closed space curve of given length with minimum total squared curvature, namely a variation of Euler's elastica which is not confined to a plane. \citet{giomi2012} accordingly referred to the coupled problem as the Euler--Plateau problem.  

A linkage between minimal surfaces and closed elastic space curves provides an interesting problem from the mathematical and physical points of view. Additionally, soap films bounded by elastic loops can serve as prototypes for various biological systems in which surface and edge energies compete. Discoidal high-density lipoprotein particles, the gut tube anchored by the dorsal mesentery membrane, and bacterial biofilms consisting of hydrated matrices of microorganisms and DNA are among such systems. In consideration of the twisted nature of the bio-filament/tube in all of these examples, \citet{biria2014} noted the importance of accounting for torsional energy. Accordingly, they considered a generalization of the Euler--Plateau problem in which the soap film spans a closed Kirchhoff rod resistant not only to bending but also to twisting about its centerline. The rod was taken to be uniform, isotropic, inextensible, and unshearable, with circular cross-section and centerline coincident with the boundary of the spanning surface. With these provisions, the rotation of a director residing in the normal cross section of the rod about the tangent to the centerline provides a useful measure of the twist density. The governing equilibrium equations were derived in a parameterized setting and a linearized buckling analysis was performed. There are, however, two shortcomings to such a treatment of the problem. First, the linear analysis does not capture the post-buckled behavior of the system. Second, since only stable solutions to the Euler--Lagrange equations are observable, some buckling modes obtained from such solutions are irrelevant.  

The aim of the present work is to build upon the previous work of \citet{biria2014} by performing a stability analysis and, further, to conduct an experimental test of the theory. Motivated by the application to DNA minicircles, a considerable literature has developed in relation to the stability of twisted rings. Rather than reviewing those works, we refer to the  discussion and citations of \citet{hoffman2004} for a comprehensive synopsis of the developments. The stability of fluid surfaces has also been studied extensively in connection with capillarity and wetting problems, as exemplified by works of \citet{vogel2000} and Rosso and \citet{rosso2004}. 

A circular disk provides a trivial solution to the Euler--Plateau problem. Considering a radial perturbation to the energy corresponding to such a solution, \citet{giomi2012} investigated the stability of a circular disk with respect to an elliptical deformation. \citet{chen2014} parameterized the spanning surface and the elastic line by a bijective mapping defined on a closed circular disk and obtained the second energy variation in terms of such a mapping. Using this approach, they studied stability of a circular disk subjected to in-plane or transverse perturbations. Here, perturbations of the angle of twist about the centerline of the cross section of the bounding loop are also allowed. The derived stability conditions are purely intrinsic, in that they do not rely on any specific shape or any particular choice of coordinates. As an illustrative example, those conditions are used to analyze the stability of a circular disk. 

Following \citet{giomi2012} and \citet{mora2012}, who considered a soap film spanning a loop of fishing line as a realization of the Euler--Plateau problem, experiments have been performed with loops of fishing line of various lengths, bending rigidities, and twist densities. This makes it possible to study the influence of salient dimensionless parameters. The effect of  twist about the centerline of the bounding loop, which has been neglected in previous experiments, is taken into account and qualitative observations are supplemented by quantitative analysis. 

The remainder of the paper is organized as follows. The necessary geometrical quantities, notation, and assumptions are introduced in Section~\ref{preliminaries}. The first and second variation conditions stemming from the energy stability criterion are summarized in Section~\ref{minima}. The stability conditions are applied to a circular disk in Section~\ref{stb}. Experimental procedures and results appear in Section~\ref{experiment}, followed by concluding remarks in Section~\ref{Conclusion}. A detailed account of the  calculations leading to the first and second variation conditions is contained in the Appendix.

\section{Preliminaries}
\label{preliminaries}

\begin{figure}[!t]
\centering
\includegraphics [width=16cm]{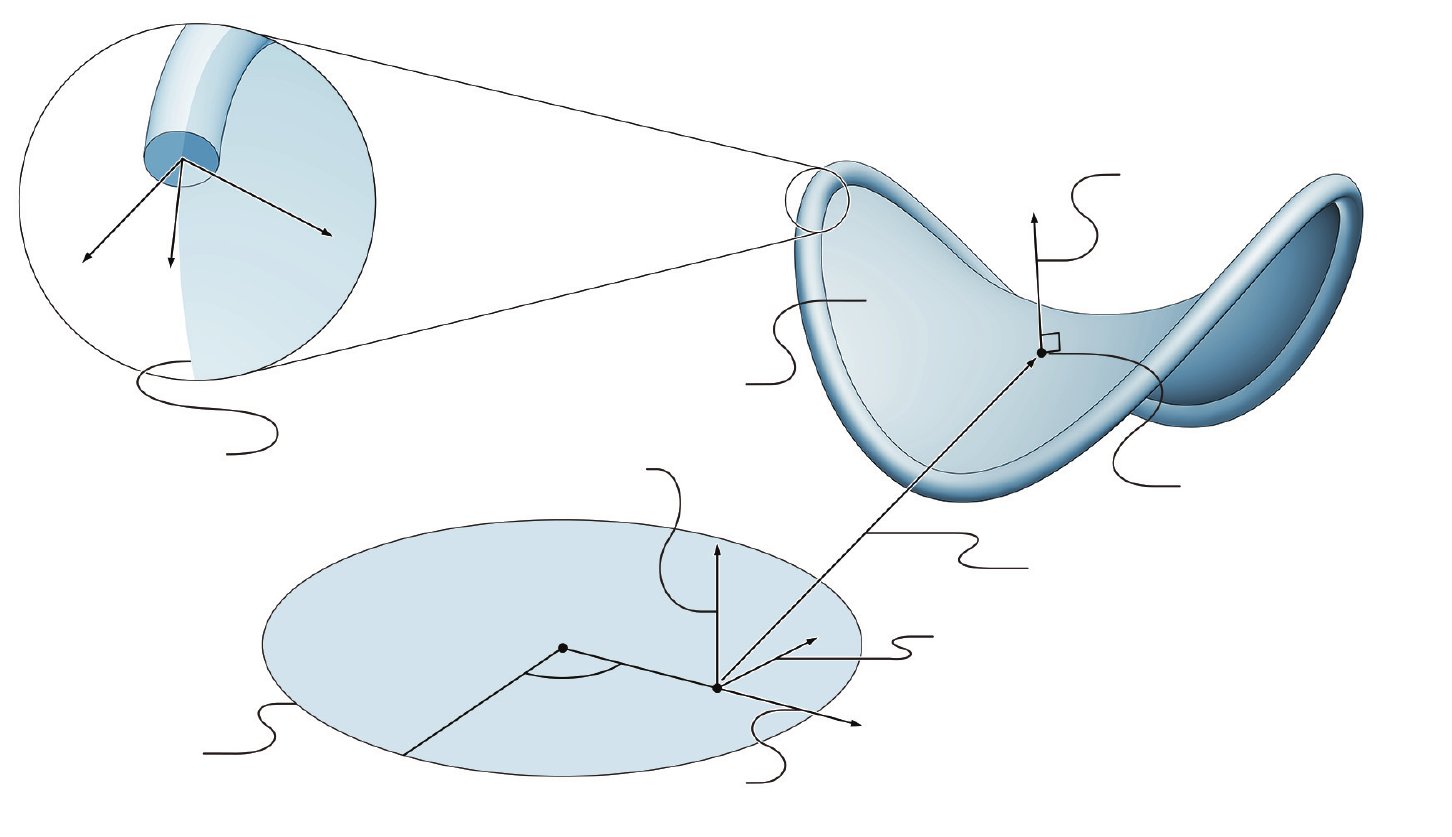}
\put(-434.5,190){\small$\bfd_1$}
\put(-362,197){\small$\bfd_2$}
\put(-396.5,182){\small$\bfd_3$}
\put(-230,139){\small$\calS$}
\put(-82,107.75){\small$\bfx=R\mskip2mu\bfxi(r,\theta)$}
\put(-130,82){\small$\bfo+R\bfxi(r,\theta)-r\bfe(\theta)$}
\put(-159,60.5){\small$\etthet$}
\put(-242,14){\small$\erthet$}
\put(-314,113){\small$\ezthet$}
%\put(-220,32){$(r,\theta)$}
\put(-277,40){\small$\theta$}
\put(-254,56.5){\small$r$}
\put(-100,205){\small$\bfm(r,\theta)$}
%origin
\put(-282,64){\small$\bfo$}
%\put(-350,162){$\bfx=R\mskip2mu\bfxi(r,\theta)$}
\put(-419,23){\small$r=1$}
\put(-423,117){\small$\calC=\partial\calS$}
%\put(-382,173){$\bfd_1$}
%\put(-310,180){$\bfd_2$}
%\put(-344,166){$\bfd_3$}
%\put(-195,125.5){$\calS$}
%\put(-60,93.5){$\bfx=R\mskip2mu\bfxi(r,\theta)$}
%\put(-98,76){$R\bfx(r,\theta)-r\bfe(\theta)$}
%\put(-129,57.5){$\et$}
%\put(-166,25){$\er$}
%\put(-248,104){$\ez$}
%\put(-220,32){$(r,\theta)$}
%\put(-98,167){$\bfm(r,\theta)$}
%%origin
%\put(-242,58){$\bfo$}
%%\put(-350,162){$\bfx=R\mskip2mu\bfxi(r,\theta)$}
%\put(-371,20){$r=1$}
%\put(-22,118){$\calC=\partial\calS$}
\caption{A soap film represented by an orientable surface $\calS=\{\bfx:\bfx=R\bfxi(r,\theta),0\le r\le1,0\le\theta\le2\pi\}$ spanning a flexible loop represented by a closed rod with centerline $\calC=\{\bfx:\bfx=R\bfxi(1,\theta),0\le\theta\le2\pi\}$ coincident with the boundary $\partial\calS$ of $\calS$. The surface $\calS$ has unit normal $\bfm$ and the curve $\calC$ is endowed with a triad $\{\bfd_1,\bfd_2,\bfd_3\}$ of orthonormal directors, the first two of which reside in the normal cross-section of the rod and the third of which is tangent to $\calC$.}
\label{f1}
\end{figure}

We consider a soap film spanning a loop of fishing line. The film is identified with a surface $\calS$ endowed with uniform tension and the loop is identified with an inextensible and unshearable elastic rod, with centerline $\calC$, that resists bending along with twisting about its tangential axis. See Figure~\ref{f1} for a schematic.

\subsection{Geometry and kinematics}

We assume that the surface $\calS$ is orientable, with unit normal field $\bfm$. The mean curvature $H$ of $\calS$ is then given by 
\be
H=-\half\divS\bfm,
\label{meanc}
\ee
where $\divS$ denotes the surface divergence on $\calS$. Further, we assume that the boundary $\partial\calS$ coincides with the centerline $\calC$ of the rod, so that
\be
\partial\calS=\calC,
\label{coincidence}
\ee
as illustrated in Figure~\ref{f1}. 

In line with the special Cosserat theory of rods, as presented, for example, by \citet{antman2005}, we endow $\calC$ with an orthonormal triad $\{\bfd_1,\bfd_2,\bfd_3\}$ of directors. For an inextensible and unshearable rod, the normal $\bfd_3$ to each cross-section coincides with the unit vector $\bft$ tangent to $\calC$, so that 
\be
\bfd_3=\bft.
\label{directors0}
\ee
Further, the remaining directors $\bfd_1$ and $\bfd_2$ lie in the plane of the Frenet unit normal $\bfn$ and unit binormal $\bfb$ to $\calC$. It follows that $\bfd_1$ and $\bfd_2$ must be determined by a single degree of freedom, namely the angle $\psi$ needed to rotate them into $\bfn$ and $\bfb$,  so that
\be
\bfd_1=(\cos\psi)\bfn+(\sin\psi)\bfb
\qquad\text{and}\qquad
\bfd_2=-(\sin\psi)\bfn+(\cos\psi)\bfb.
\label{directors}
\ee

The curvature $\kappa$ and torsion $\tau$ of $\calC$ are defined in accord with %the relations
\be
\kappa=|\bft'|
\qquad\text{and}\qquad
\kappa^2\tau=(\bft \times\bft^{\prime})\cdot\bft^{\prime\prime},
\label{kappatau}
\ee
where a prime denotes differentiation with respect to the arclength. By the Frenet--Serret relations
\be
\bft'=\kappa\bfn,
\qquad
\bfn'=-\kappa\bft+\tau\bfb, 
\qquad
\bfb'=-\tau\bfn,
\label{SF}
\ee
and \eqref{directors}, the strain vector $\bsu$ defined by  
\be
\bfd'_i=\bsu \times \bfd_i,  \qquad i=1,2,3,
\label{strain0}
\ee
takes the particular form 
\be
\bsu=\kappa((\sin\psi)\bfd_1+(\cos\psi)\bfd_2)+(\tau+\psi')\bfd_3.
\label{strain}
\ee
Using \eqref{directors} in \eqref{strain} and introducing the twist density
\be
\Omega=\bsu\cdot\bfd_3=\tau+\psi',
\label{twist}
\ee
leads to a simplified representation
\be
\bsu=\Omega\bft+\kappa\bfb
\label{strainvector}
\ee
for $\bsu$ relative to the tangential and binormal elements of the Frenet frame.

In terms of the normal and binormal elements $\bfn$ and $\bfb$ of the Frenet frame of $\calC$, the restriction of $\bfm$ to $\calC$ and the associated unit tangent-normal vector $\bfnu=\bft\times\bfm$ on $\calC$ admit decompositions with the form
\be
\bfm=(\cos\vartheta)\bfn+(\sin\vartheta)\bfb
\qquad\text{and}\qquad
\bfnu=-(\sin\vartheta)\bfn+(\cos\vartheta)\bfb,
\label{vartheta}
\ee
where $\vartheta$ is the contact angle considered by \citet{giomi2012}. Another quantity of interest defined on $\calC$ is the curvature vector
\be
\bfk=\kappa\bfn.
\label{curvec}
\ee
%
%%%%%%%%%%%%%%%%%%%
\subsection{Constitutive assumptions}
The soap film is assumed to have uniform surface tension
\be
\sigma>0.
\label{sigmagt0}
\ee

Aside from being inextensible and unshearable, the bounding loop is assumed to be isotropic and hyperelastic, with strain energy-density function $W$ quadratic in the components of the strain vector $\bsu$ defined in \eqref{strainvector} but of the particular form
\be
W=\sum^3_{i=1} \half k_i \go \bsu \cdot \bfd_i,
\qquad
\left\{ 
\begin{split}
&k_1=k_2=a>0,
\\[4pt]
&k_3=c>0,
\end{split}
\,\right.
\label{strain energy}
\ee
where $a>0$ and $c>0$ respectively represent the uniform bending rigidity and uniform torsional rigidity of the bounding loop. Granted \eqref{strain energy} and that gravitational effects are negligible, the net free-energy $E$ of the system takes the form   
\be
E=
\int_{\calC}\half(a\kappa^2+\crig \Omega^2)+\int_{\calS} \sigma.
\label{EL1}
\ee
The first and second terms on the right-hand side of \eqref{EL1} are referred to as the lineal and areal contributions to $E$, respectively. 

A circular loop with length $2\pi R$ and uniform twist density $\Omega$ that is spanned by a flat circular disk can be regarded as a ground state of the system. For any such ground state, \eqref{EL1} specializes to yield an expression
\be
\frac{R E}{\pi a}=1+\frac{a}{c}\Big(\frac{R\Omega c}{a}\Big)^{\!\!2}+\frac{R^3\sigma}{a}
\label{Eground}
\ee
for the ratio of the net free-energy $E$ of the system to the lineal bending-energy $\pi a/R$. Inspection of \eqref{Eground} reveals the potential significance of three dimensionless parameters:
\be
\chi=\frac{c}{a}>0,
\qquad
\mu=\frac{R\Omega c}{a},
\qquad\text{and}\qquad
\eta=\frac{R^3\sigma}{a}>0.
\label{dgroups1}
\ee
While $\chi$ measures the importance of the torsional rigidity $c$ relative to that of the bending rigidity $a$, $\mu$ combines $\chi$ with the dimensionless twist density $R\Omega$. Further, $\eta$ measures the importance of the areal free-energy $\pi R^2\sigma$ relative to that of the lineal bending-energy $\pi a/R$ and in the absence of twist is the only dimensionless parameter of consequence. Notice that, in contrast to $\chi$ and $\eta$, which must be positive by \eqref{sigmagt0} and \eqref{strain energy}, the sign of $\mu$ is unrestricted.

%%%%%%%%%%%%%%%%%%%%%%%%%%%%%%%%%%%%
\section{Necessary and sufficient conditions for energy minima}
\label{minima}

In the present context, the energy stability criterion, as treated by \citet{ericksen1966thermo}, states that a stable equilibrium of the system consisting of the bounding loop and the spanning surface is given by a parameterization that minimizes the net free-energy $E$ subject to the constraint that $\calC$ be inextensible. This requirement, which amounts to satisfying the first and second variation conditions
\be
\delta E=0
\qquad\text{and}\qquad
\delta^2E\ge0
\label{fsvc}
\ee
for all admissible variations, should not be confused with notions of dynamical stability.

\subsection{First variation condition}
\label{First variation condition}

Let $\bfr$ provide an arclength parameterization of $\calC$, so that $\bfr'=\bft$. Given variations
\be
\bfv=\delta\bfr
\qquad\text{and}\qquad
\iota=\delta\psi
\label{vandiota}
\ee
of $\bfr$ and the twist angle $\psi$, the first variation $\delta F$ of $F$ is
\be
\delta E=\int_{\calC}(a\kappa\del\kappa+\crig\Omega\del\Omega
+\gamma \go \bft\cdot\bfv'\mskip1mu)+ \del \int_{\calS} \sigma,
\label{EL2}
\ee
where $\gamma$ is a Lagrange multiplier  needed to ensure the inextensibility of $\calC$. With the expressions \eqref{ap10}, \eqref{vark}, and \eqref{varomega2} for $\del\int_{\calS}\sigma$, $\kappa\delta\kappa$, and $\Omega\delta\Omega$ provided in the Appendix, an integration by parts, and (since $\calC$ is closed) the periodicity of $\kappa$, $\Omega$, $\gamma$, and $\iota$, \eqref{EL2} becomes 
%
%\begin{multline}
\be
\delta E=
\int_{\calC}((a\bfk'
-\crig\Omega \kappa \bfb-c(\kappa^{-1}\Omega'\bfb)'
-\gamma\bft)'+\sigma\bfnu)\cdot\bfv
-\int_{\calC}\crig\Omega'\mskip1.5mu\iota
-\int_{\calS}2\sigma H\bfm\cdot\bfv.
\label{EL4}
\ee

If $\bfv$ has compact support about a point interior to $\calS$ and $\iota$ vanishes on $\calC$, then applying the first variation condition \eqref{fsvc}$_1$ to \eqref{EL4} yields the areal equilibrium condition,
\be
H=0,
\label{mean}
\ee
familiar from the classical Plateau problem. Physically, condition \eqref{mean} represents the component of force balance on $\calS$ in the direction normal to $\calS$.

With \eqref{mean}, the areal contribution to \eqref{EL4} vanishes, reducing it to
\be
\delta E=
\int_{\calC}((a\bfk'-\crig\Omega \kappa \bfb-c(\kappa^{-1}\Omega'\bfb)'-\gamma\bft)'
+\sigma\bfnu)\cdot\bfv
-\int_{\calC}\crig\Omega'\mskip1.5mu\iota.
\label{EL4reduced}
\ee
The restriction of $\bfv$ to $\calC$ and $\iota$ may be chosen independently. If, in particular, the restriction of $\bfv$ to $\calC$ vanishes and $\iota$ has compact support about a point on $\calC$, then applying the first variation condition \eqref{fsvc}$_1$ to \eqref{EL4reduced} yields a lineal equilibrium condition,
\be
\Omega'=0,
\label{EL6}
\ee
which represents the tangential component of moment balance on $\calC$.

Finally, if the restriction of $\bfv$ to $\calC$ has compact support about a point on $\calC$, then applying the first variation condition \eqref{fsvc}$_1$ to the reduction
\be
\delta E=-\int_{\calC}((a\bfk'-\crig\Omega \kappa \bfb-\gamma\bft)'+\sigma\bfnu)\cdot\bfv
\label{EL4reducedagain}
\ee
of \eqref{EL4reduced} implied by \eqref{EL6} yields an additional lineal equilibrium condition,
\be
(a\bfk'-\crig\Omega \kappa \bfb-\gamma\bft)'+\sigma\bfnu=\bf0,
\label{EL7}
\ee
which represents force balance on $\calC$. 

From \eqref{EL7}, $-a\bfk'+\crig\Omega\kappa\bfb-\gamma\bft$ is the distributed resultant internal force of $\calC$ and $-\sigma\bfnu$ is the distributed external force, per unit length, exerted by $\calS$ on $\calC$. By the Frenet--Serret relations \eqref{SF} and the definition \eqref{curvec} of the curvature vector $\bfk$, the resultant internal force of $\calC$ can be expressed as
\be
-a\bfk'+\crig\Omega \kappa \bfb
+\gamma\bft=
(a\kappa^2+\gamma)\bft
-a\kappa'\bfn-(a\tau-\crig\Omega)\kappa\bfb,
\label{rif}
\ee
from which it is evident that twist generates an internal force binormal to $\calC$. 

In view of the Frenet--Serret relations \eqref{SF} and the representation \eqref{vartheta}$_2$ of the tangent-normal vector $\bfnu$, and the moment equilibrium condition \eqref{EL6}, isolating the components of \eqref{EL7} in the directions tangent, normal, and binormal to $\calC$ results in a system of three scalar conditions,
\be
\left.
\begin{array}{c}
\displaystyle
\gamma+\frac{3}{2}a\kappa^2=\beta,
%\text{constant} 
\cr\noalign{\vskip8pt}
\displaystyle
\kappa''+\half \kappa^3-\kappa\Big(\tau^2-\frac{\crig\Omega}{a}
\tau
+\frac{\beta}{a}\Big)-\frac{\sigma}{a}\sin{\vartheta}=0,
\cr\noalign{\vskip8pt}
\displaystyle
2\kappa'\Big(\tau-\frac{\crig\Omega}{2a}\Big)+\kappa\tau'+\frac{\sigma}{a}\cos{\vartheta}=0,
\end{array}
\right\}
\label{EL8}
\ee
wherein $\beta$ is a constant and $\vartheta$ is the contact angle defined in \eqref{vartheta}. The equilibrium conditions \eqref{EL8} coincide with equations derived previously by \citet{biria2014} as a consequence of a vectorial equilibrium condition arising in the context of a parameterized description akin to that utilized in Section~\ref{stb}. For $\sigma=0$, the external force acting on the boundary vanishes and, hence, \eqref{EL8}$_2$ and \eqref{EL8}$_3$ specialize to the Euler--Lagrange equations for a Kirchhoff elastic rod, as developed by \citet{ls}. Additionally, for the degenerate case $c=0$,  \eqref{EL8}$_2$ and \eqref{EL8}$_3$ reduce to the equilibrium equations obtained by \citet{giomi2012}.
Finally, notice that $\vartheta=\pi/2$ and $\tau=0$ for a planar configuration, in which case \eqref{EL8}$_3$ is satisfied trivially and \eqref{EL8}$_2$ reduces to
\be
\kappa''+\frac{1}{2}\kappa^3- \frac{\beta}{a}\kappa-\frac{\sigma}{a}=0,
\label{planar}
\ee
which is equivalent to an equation governing the equilibrium of a uniformly pressurized elastic tube considered by \citet{antman1968}. Analytical solutions of equations equivalent to \eqref{planar} have been obtained by \citet{watanabe2008} and \citet{djondjorov2011}.

%%%%%%%%%%%%%%%%%%%%%%
\subsection{Second variation condition}
\label{stability}

In view of the expression \eqref{EL4reduced} for the first variation $\del E$ of $E$, the identities \eqref{ap11}, \eqref{inex1}, \eqref{var2k2}, and \eqref{var2omega2} established in the Appendix yield
\begin{multline}
\del^2 E=
\int_{\calC} (a \abs{ \bfv''}^2+\gamma \abs{\bfv'}^2+ c((\kappa^{-1} \bfb \cdot \bfv'')'+\kappa \go \bfb \cdot \bfv'+\iota')^2+\crig \Omega(\bfv'' \times \bfv')\cdot \bft)
\\[4pt]
 -\int_{\calC}\Omega' (\del(\kappa^{-1}\bfb\cdot\bfv'')+\kappa\bfb\cdot\tilde{\bfv}+\del \iota )+ \int_{\calC}(a \bfk'' - \crig \Omega (\kappa \bfb)'- \gamma \bfk +\sigma \bfnu) \cdot  \tilde\bfv 
\\[4pt]
+\int_{\calS} \sigma((\divs\bfv)^2-\tr((\grads\bfv)^2)
+\abs{(\grads{\bfv})^{\trans}\bfm}^2-2H\tilde \bfv \cdot \bfm).
\label{EL9}
\end{multline}
Using the equilibrium conditions \eqref{mean}, \eqref{EL6}, and \eqref{EL7} to reduce \eqref{EL9} and invoking an additional identity \eqref{var2omega3} from the Appendix yields a simplified expression for $\del^2E$, and delivers the second variation condition \eqref{fsvc}$_2$ in the form
\begin{multline}
\del^2E=\int_{\calC}(\abs{ \bfv''}^2+\frac{\gamma}{a} \abs{\bfv'}^2+ \mu (\bfv'' \times \bfv')\cdot \bft
+\chi((\kappa^{-1} \bfb \cdot \bfv'')'+\kappa \go \bfb \cdot \bfv')^2
-\chi(\iota')^2\mskip1mu)
\\[4pt]
+\int_{\calS}\eta((\divs\bfv)^2-\tr((\grads\bfv)^2)
+\abs{(\grads{\bfv})^{\trans}\bfm}^2\mskip1mu)\geq 0.
\label{EL11}
\end{multline}

%%%%%%%%%%%%%%%%%%%%%%%%%%%%%%%%%%%%%%%%%%%%
\section{Stability analysis}
\label{stb}
We suppose that $\calS$ can be parameterized in accord with
\be
\calS=\{\bfx:\bfx=R\bfxi(r,\theta), 0\le r\le1, 0\le \theta \le 2\pi\},
\label{Srep}
\ee
where $R$ is the radius of a circle with perimeter equal to the length of $\calC$ and $\bfxi$ is a four-times continuously differentiable, injective mapping defined on the closed unit disk. As a consequence of the assumption \eqref{coincidence} that $\partial\calS$ and $\calC$ coincide, the surface parameterization \eqref{Srep} induces a parameterization
\be
\calC=\{\bfx:\bfx=R\bfxi(1,\theta),0\le \theta \le 2\pi\}
\label{Crep}
\ee
of $\calC$. Due to the lineal inextensibility constraint, the restriction of $\xit$ to $r=1$ must satisfy
\be
|\xit|_{\reqone}=1.
\label{Pr1}
\ee
Granted \eqref{Crep} and \eqref{Pr1}, the curvature, torsion, and twist denisty of $\calC$ admit representations
\be
\kappa=\frac{|\xitt|}{R}\Big|_{\reqone},
\qquad 
\tau=\frac{(\xit\times\xitt)\cdot\xittt}{R|\xitt|^2}\Big|_{\reqone}, 
\qquad\text{and}\qquad
\Omega=\tau+\frac{\psi_\theta}{R}.
\label{kappatauomegaparrep}
\ee

On introducing the dimensionless free-energy $\Phi={R E}/{a}$ and the variation
\be
\bfu=\del\bfxi
\ee
of $\bfxi$, \eqref{EL1} yields
\be
\Phi=
\lint \frac{1}{2} \Big[|\xitt|^2+\chi\Big(\frac{(\xit\times\xitt)\cdot\xittt}{|\xitt|^2}
+\psi_\theta\Big)^{\!\!2}\,\Big]_{\reqone}\!\!\dtheta
+\aintdim\eta|\xir\times\xit|\drdtheta.
\label{Pr2}
\ee
Furthermore, the parameterized versions of the first variation \eqref{EL4reduced} and the second-variation condition \eqref{EL11} follow as
\begin{multline}
\del \Phi
=\lint\Big[\eta(\xit \times \bfm)\cdot \bfu
-\Big(\xittt-\mu \frac{ \xitt \times \xittt}{|\xitt|^2}-\gamma \xit \Big) \cdot \ut \Big]_{\reqone}\dtheta
\\[4pt]
+\lint \chi \Big[\Big( \Big(\frac{(\bfxi_{\theta}\times\bfxi_{\theta\theta})\cdot\bfxi_{\theta\theta\theta}}{|\bfxi_{\theta\theta}|^2} +\psi_\theta \Big)_{\!\!\theta}\frac {\xit \times \xitt}{|\xitt|^2}\Big)_{\!\!\theta} \cdot  \ut
-\Big(\frac{(\bfxi_{\theta}\times\bfxi_{\theta\theta})\cdot\bfxi_{\theta\theta\theta}}{|\bfxi_{\theta\theta}|^2}
+\psi_\theta
\Big)_{\!\!\theta}\iota \Big]_{\reqone}\dtheta
\\[8pt]
-\aintdim\eta(\xit \times \bfm_\rho+\bfm_\theta \times \xir) \cdot \bfu\drdtheta,
\label{nd1}
\end{multline}
and
\begin{multline} 
\del^2\Phi=   
\lint\Big[|\utt|^2 +{\mu} (\utt \times \ut) \cdot \xit
+\gamma |\ut|^2 
 \\[4pt]
 + \chi \Big( \frac{( \xitt \times \xittt)\cdot \ut}{|\xitt|^2}+\Big( \frac {(\xit \times \xitt) \cdot \utt} {|\xitt|^2} \Big)_{\!\theta}\,\Big)^2 - \chi \go \iota_{\theta}^2 
\Big]_{\reqone}\dtheta
\\[4pt]
+\aintdim \eta \Big(\frac{|\bfP(\bfu_r \times \xit+\xir \times \ut)|^2}{|\xir \times \xit|}+2(\bfu_r \times \ut)\cdot\bfm\Big)\drdtheta\geq0.
\label{nd2}
\end{multline}
On setting $\chi=0$ (and consequently, by \eqref{dgroups1}$_2$, $\mu=0$), the stability requirement \eqref{nd2} reduces to that of the Euler--Plateau problem, as presented by \citet{chen2014}. The areal term involves the first partial derivatives of both $\bfxi$ and $\bfu$. However,  the partial derivatives of $\bfxi$ are incorporated in the lineal terms only if twist is taken into account. On neglecting torsional effects by setting $\chi=0$, only partial derivatives of the variation $\bfu$ remain in the lineal contribution to \eqref{nd2}. This implies that the influence of the ground state at the boundary is stronger if twisting energy is taken into account. 

Let $\er$ and $\et$ denote the radial and azimuthal basis vectors associated with the coordinates $r$ and $\theta$, as depicted in Figure~\ref{f1}. \citet{biria2014} showed that for any given constant value of $\psi_\theta$, a planar disk bounded by a circle of unit radius, with $\bfxi$ and $\gamma$ given by
\be
\bfxi(r, \theta)=r\er
\qquad\text{and}\qquad
\gamma=-(1+\eta),
\label{ts1}
\ee
renders the right hand of \eqref{nd1} zero and, thus, provides a trivial solution to the Euler--Lagrange equations. For a generic perturbation
\be
\bfu=v \er+u \et+w \ez
\label{tss2}
\ee
of the trivial solution \eqref{ts1} subject to the consequences
\be
[\ut\cdot\et]_{\reqone}=[\uit+v]_{\reqone}=0
\label{ts3}
\ee
arising from the constraint of inextensibility, the stability condition \eqref{nd2} requires that
\be
\iota_\theta=0
\label{ts7}
\ee
and that
\begin{multline}
\int_0^{2\pi}[\witt^2-(1+\eta+\mu^2)\wit^2 +(\vitt+v-{\mu}\wit)^2-\eta(\vit^2-v^2)]_{\reqone}\dtheta
\\[4pt]
+\aintdim\eta\Big(r\wir^2+\frac{1}{r}\wit^2\Big)\drdtheta\ge0.
\label{ts8}
\end{multline}
Recall that, in the absence of torsion, the equilibrium equation \eqref{EL6} implies that the twist angle changes uniformly along $\calC$. The stability requirement \eqref{ts7} further guarantees the preservation of this uniformity after buckling to another stable equilibrium configuration. As a consequence of \eqref{ts8}, the radial and transverse perturbations $v$ and $w$ are coupled through the parameter $\mu$. Purely radial and purely transverse perturbations, for which \eqref{ts8} reduces to the stability condition investigated by \citet{chen2014}, are therefore indifferent to the dimensionless parameter $\mu$ associated with twist.

The radial and transverse perturbation components $w$ and $v$ admit Fourier expansions 
\be
\left.
\begin{split}
w(r,\theta)=a_0(r)+ \sum_{n=1}^{\infty} a_n(r) \cos n\theta + b_n(r) \sin n\theta,
\\[4pt]
v(r,\theta)=c_0(r)+ \sum_{n=1}^{\infty} c_n(r) \cos n\theta + d_n(r) \sin n\theta,
\label{F1}
\end{split}
\mskip4mu\right\}
\ee
which when substituted into \eqref{ts8}$_2$ gives
\begin{multline}
%\del^2\Phi=
2\pi\eta\int_0^{1}(a'_0(\rho))^2r\dr 
%\\[4pt]
+\pi\eta\sum_{n=1}^{\infty} \int_0^{1}\Big((a'_n(r))^2+(b'_n(\rho))^2+
\frac{n^2}{r^2}(a^2_n(r)+b^2_n(\rho))\Big)r\dr 
\\[4pt]
+\pi \sum_{n=1}^{\infty}\Big((n^4-(\eta+1)n^2)(a^2_n(1)+b^2_n(1))- (n^2-1)(\eta-n^2+1)(d^2_n(1)+c^2_n(1))\Big)
\\[4pt]
+\pi \sum_{n=1}^{\infty}\Big(2\mu n(n^2-1)(a_n(1)d_n(1)-c_n(1)b_n(1))\Big)\geq 0.
\label{F2}
\end{multline}
The second term on the left-hand side of \eqref{F2} can be rewritten as
\begin{multline}
\pi\eta\sum_{n=1}^{\infty} \int_0^{1}\Big((a'_n(r))^2+(b'_n(r))^2+\frac{n^2}{r^2}(a^2_n(r)+b^2_n(r)) \big\}r\dr 
\\[4pt]
=\pi\eta\sum_{n=1}^{\infty} \int_0^{1}\Big(\Big(a'_n(r)-\frac{n}{r} a_n(r)\Big)^{\!\!2}
+\Big(b'_n(r)-\frac{n}{\rho} b_n(r)\Big)^{\!\!2}\,\Big)r\dr 
\\[4pt]
+ \pi\eta\sum_{n=1}^{\infty} n(a^2_n(1)-a^2_n(0)+b^2_n(1)-b^2_n(0)),
\end{multline}
which, since the integrability of $(a^2_n(r)+b^2_n(r))/r^2$ implies that $a_n(0)=b_n(0)=0$, leads to the alternative stability condition
\begin{multline}
2\pi\eta\int_0^{1}(a'_0(\rho))^2r\dr 
%\\[4pt]
+\pi\eta\sum_{n=1}^{\infty} \int_0^{1} \Big(\Big(a'_n(r)-\frac{n}{r} a_n(r)\Big)^{\!\!2}
+\Big(b'_n(r)-\frac{n}{r}b_n(r)\Big)^{\!\!2}\,\Big)r\dr 
\\[4pt]
+\pi \sum_{n=1}^{\infty}\Big((n^4-(\eta+1)n^2+\eta n)(a^2_n(1)+b^2_n(1))
- n^2-1)(\eta-n^2+1) (d^2_n(1)+c^2_n(1))\Big)
\\[4pt]
+\pi \sum_{n=1}^{\infty}  \Big(2\mu n(n^2-1) (a_n(1)d_n(1)-c_n(1)b_n(1))\Big)\geq 0.
\label{F3}
\end{multline}
Since $\eta$ is positive, the first two terms on the left-hand side of \eqref{F3} are always nonnegative. To satisfy \eqref{F3}, it is therefore sufficient to satisfy the inequalities
\be
(n^4-(\eta+1)n^2+\eta n)a^2_n(1)- (n^2-1)(\eta-n^2+1)d^2_n(1)+2\mu n(n^2-1) a_n(1)d_n(1) \geq 0
\label{F4}
\ee
and 
\be
(n^4-(\eta+1)n^2+\eta n)b^2_n(1)- (n^2-1)(\eta-n^2+1)c^2_n(1)-2\mu n(n^2-1) b_n(1)c_n(1) \geq 0
\label{F5}
\ee
for each $n \geq 1$. The left-hand sides of \eqref{F4} and \eqref{F5} are quadratic forms corresponding to matrices that share the same trace and determinant. These quadratic forms are positive-semidefinite if
\be
(n^2-1)(\eta-n^2+1) \le 0
\ee
and
\be
\mu^2n^2(n^2-1)^2-(n^4-(\eta+1)n^2+\eta n)(n^2-1)(\eta-n^2+1) \leq 0
\ee 
or, equivalently, since $n\ge1$, if 
\be
\eta\le n^2-1
\qquad\text{and}\qquad
n(n-1)(n^2-1)p(\eta;n,\mu)\ge0,
\label{F7}
\ee 
with
\be
p(\eta;n,\mu)=\eta^2-(2n-1)(n+1)\eta-n(n+1)(\mu^2-n^2+1).
\ee
Since $n\ge1$, the sign of left-hand side of \eqref{F7}$_2$ hinges on the sign of $p(\eta;n,\mu)$, which depends quadratically on $\eta$. Since $\eta$ and the discriminant of the relevant quadratic are both positive, $p(\eta;n,\mu)\ge0$ if 
\be
\eta\leq\eta_- \qquad \text{or} \qquad \eta \geq \eta_+,
\label{F8}
\ee
where the roots $\eta_+$ and $\eta_-$ of $p(\eta;n,\mu)$ are given by
\be
\eta_{\pm}= (n+1)\Big(n-\frac{1}{2}\pm \frac{1}{2}\sqrt{1+\frac{4n\mu^2}{n+1}}\,\Big).
\label{F9}
\ee
Since $\eta_{-} \leq n^2-1$ and $\eta_{+} \geq n^2-1$, the intersection of conditions \eqref{F7}$_1$ and \eqref{F8} gives the stability condition
\be
\eta \leq \eta_{-}.
\label{F10}
\ee
Since $\eta$ must be positive, \eqref{F10} holds only if $\eta_{-} \geq 0$, or equivalently only if 
\be
\mu^2 \leq n^2-1.
\label{F11}
\ee
To summarize, the circular disk is a stable equilibrium solution if $\mu$, $\eta$, and $n$ satisfy
\be
\mu^2 \leq n^2-1 
\qquad\text{and}\qquad
\eta \leq (n+1)\Big(n-\frac{1}{2}- \frac{1}{2}\sqrt{1+\frac{4n\mu^2}{n+1}}\,\Big).
\label{F12}
\ee

\begin{figure}[t]
\centering
%\vspace{-8pt}
\includegraphics [width=4 in]{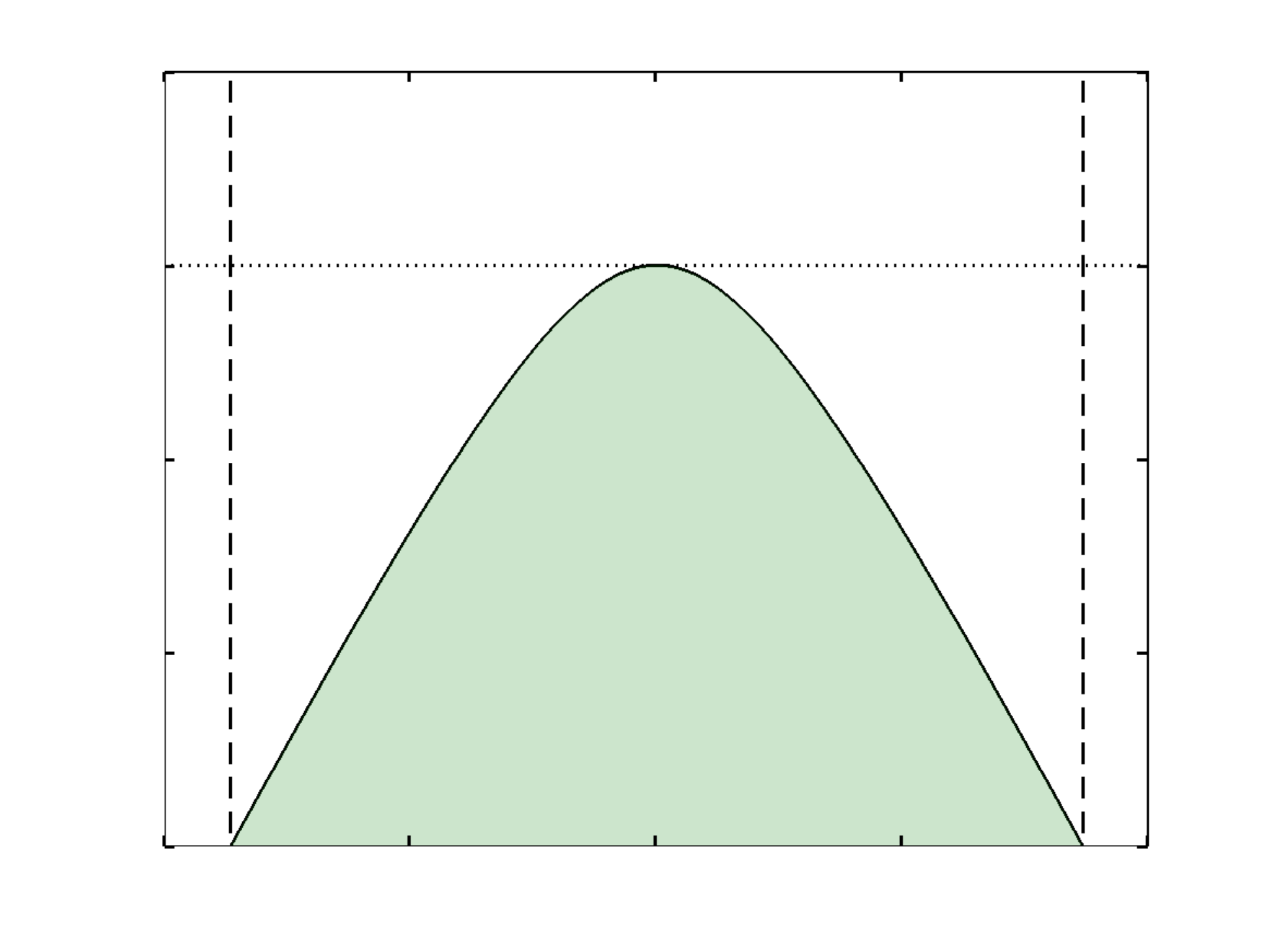}
\put(-286,109){$\eta$}
\put(-142,-6){$\mu$}
\put(-260,197){$4$}
\put(-260,153){$3$}
\put(-260,109){$2$}
\put(-260,64){$1$}
\put(-260, 20){$0$}
\put(-254,10){$-\sqrt{3}$}
\put(-198,10){$-1$}
\put(-142,10){$0$}
\put(-86,10){$1$}
\put(-55,10){$\sqrt{3}$}
%\vspace{-5pt}
\caption{The region of stability: If the bounding loop is not spanned by a soap film (or, formally, the surface tension $\sigma$ is required to vanish), the inequalities \eqref{F12} reduce to the \citeapos{Michell} condition $-\sqrt{3} \leq \mu \leq \sqrt{3}$ for the stability of a twisted ring. If the bounding loop does not resist twisting about its centerline (or, formally, the torsional rigidity is required to vanish), the inequalities \eqref{F12} reduce to the condition $0 <\eta \leq 3$ for the stability of a circular disk obtained by \citet{chen2014}.}
\label{f2}
\end{figure}
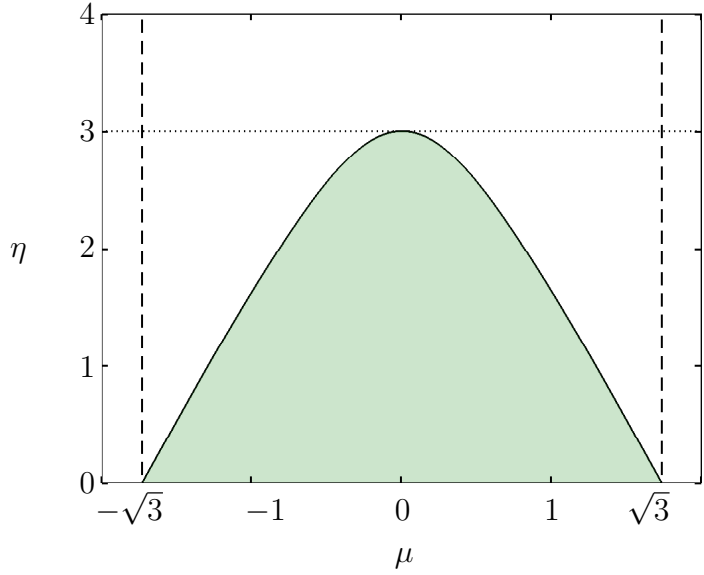
The first mode that becomes unstable is $n=2$, for which the corresponding stability region is demonstrated in Figure~\ref{f2}. For $\eta=0$, the inequalities \eqref{F12} reduce to the single condition $|\mu|<\sqrt{3}$, which agrees with \citeapos{Michell} condition for the stability of a twisted ring. Moreover, for $\mu=0$, the inequalities \eqref{F12} reduce to the single condition $\eta<3$, which agrees with the stability condition for the Euler--Plateau problem derived by \citet{chen2014}. Twisting the bounding loop destabilizes an otherwise stable circular disk, leading to buckling at values of $\eta$ below the critical value $\eta=3$ for the Euler--Plateau problem. Additionally, the force exerted by surface tension makes a circular loop less stable with respect to twist, leading to buckling at values of $|\mu|$ below the critical value $|\mu|=\sqrt{3}$ for a twisted ring. 

\section{Comparison with experiments}
\label{experiment}
\subsection{Eperimental procedure}
\label{experiment1}
Experiments with elastic loops spanned by soap films were performed to assess the validity of the theoretical predictions. A soap solution was made by mixing 245~mL distilled water, 5~mL of Dawn\textsuperscript{\textregistered} dish washing liquid, and 45~mL Glycerin. Using the pendant drop method with a Biolin Scientific Attension Theta optical tensiometer, the surface tension $\sigma$ of the solution was measured to be $\sigma=24.0 \pm 0.1$~mN/m. PVDF fluoropolymer fishing line with cross-sectional diameters of 0.178~mm, 0.229~mm, and 0.279~mm was used in our experiments. \citet{PVDF2} have reported values of the Poisson ratio of PVDF fluoropolymer in the range $0.33$--$0.35$. Various values for the Young's modulus of  PVDF fluoropolymer are available: \citet{PVDF1} reported a value of $1.52$~GPa; \citet{PVDF2} reported a value of $1.87$~GPa; \citet{PVDF3} reported a value of  $2$~GPa; and \citet{PVDF4} reported values between $2.5$~GPa and $2.7$~GPa. An Instron 5566 Universal Testing Machine was used to independently measure the Young's modulus of the fishing line used in our experiments. Samples of cross-sectional diameter 0.229~mm were held fixed between the grips with a gauge length of $20$~cm and were stretched at a loading rate of $200$~mN min$^{-1}$. Given the initial length of the fishing line, the strain data were constructed from the extension data. Also, using the measured cross-sectional diameter of the fishing line and approximating its cross section as uniform and circular, the stress data was generated from the load data (see Table~1). Finally, the Young's modulus $E$ that resulted from the slope of the linear region of the stress-strain diagrams was found to be $2.29 \pm 0.04$~GPa, which falls within the range of reported values cited above. 

\begin{table}[!t]%\footnotesize
%\caption{Tensile-test data used for measurement of the Young's modulus}
%\vspace{-12pt}
%\resizebox{\textwidth}{!}{%
\begin{center}
\begin{tabular}{| l || l | c | c | c | c | c |}
\hline
Load (N) & 1.05   &  2.08     &  2.89 &  3.67 & 4.32 & 4.98 \\ \hline
Extension (mm) & 1.67 &  3.33  &  5.00  &  6.67   &  8.33  & 10.0 \\ \hline
Tensile stress ($\text{MPa}\times10^{-1}$) & 2.56 & 5.06 & 7.01 & 8.90 & 10.5 & 12.1 \\ \hline
Tensile strain (\%) & 0.833 & 1.67 & 2.50 & 3.33 & 4.17 & 5.00 \\ \hline
\end{tabular}
\caption{Tensile-test data used for measurement of the Young's modulus}
\end{center}
%}
\label{table1}
%\normalsize
\end{table}

Segments of fishing-line with given lengths were held fixed at one end with an alligator clip and their free ends were twisted using a hand-drill, which was fixed on a protractor (see Figure~\ref{f2}).
\begin{figure}
\centering
\includegraphics [width=0.8\textwidth]{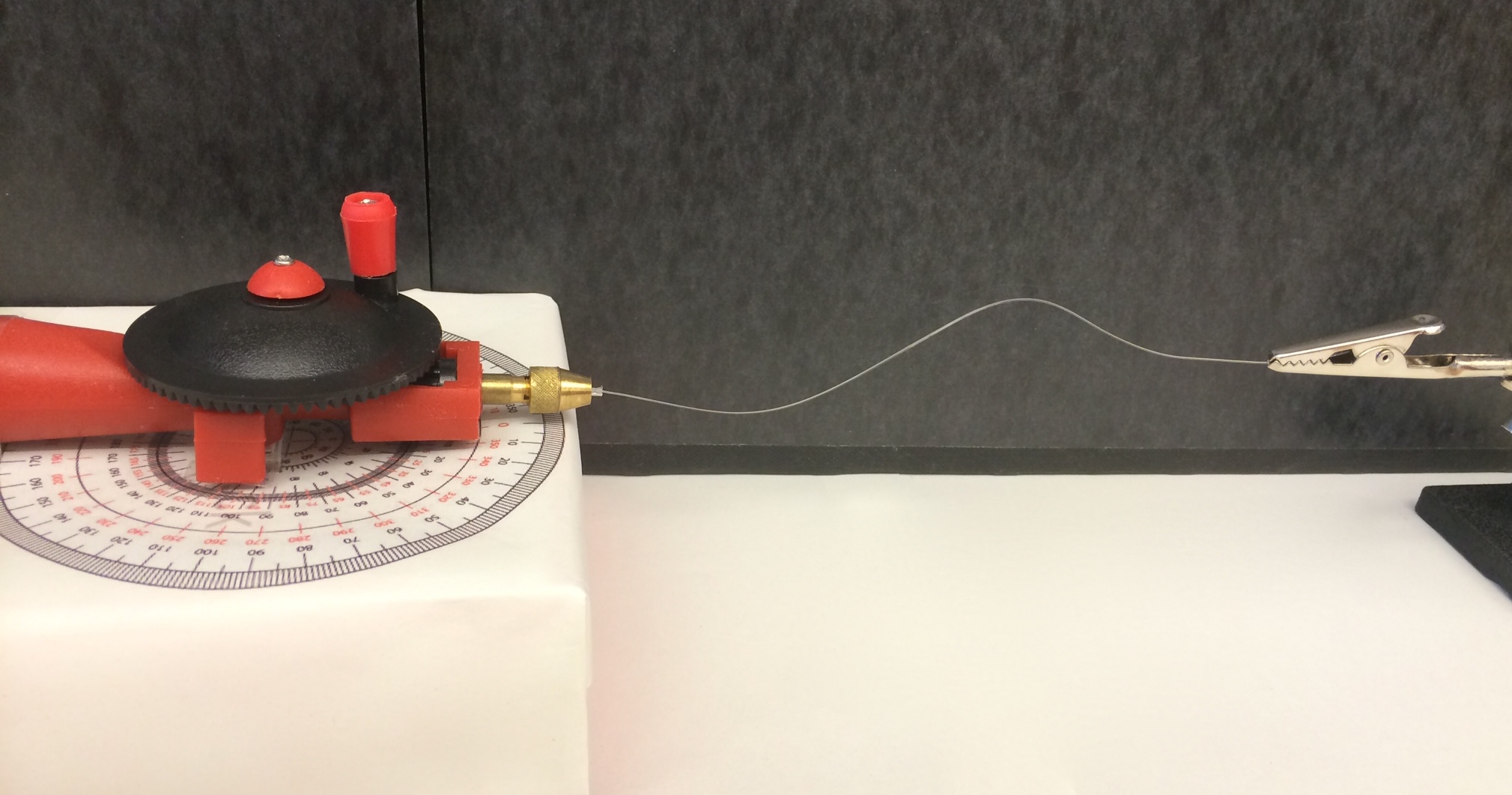}
\caption{Set-up used for measuring the twisting angle given to the fishing line prior to joining the ends to form a loop.
}
\label{f2}
\end{figure}
The total angle of twist, denoted here by $\Psi$, was read from the protractor and subsequently the ends of the segments were glued together. This procedure allowed for the fabrication of loops of given length $L=2\pi R$ and twist density $\Omega=\Psi/L$. The number of teeth of the driver gear of the hand drill used were greater by a factor of three than those of the driven gear, which results in an absolute error of $3^\circ$ for the measurement of the total angle of twist. 

For different net twisting angles, loops of lengths varying by increments of $0.5$~cm were made and dipped into and removed from the solution. The size at which the loop buckled out of plane from a circle was recorded. Through this procedure, we determined the critical radius at which buckling occurs to within an error of 1~mm. 

To facilitate visualization under UV light and capture images of representative configurations, we added Fluorescein to the soap solution. Measurements confirmed that this did not tangibly alter the surface tension of the soap solution.

%%%%%%%%%%%%%%%%%%%%%%%%%%%%%
\subsection{Experimental results}
\label{experiment2}

\begin{figure}[t]
\centering
{\includegraphics[width=0.2125\textwidth]{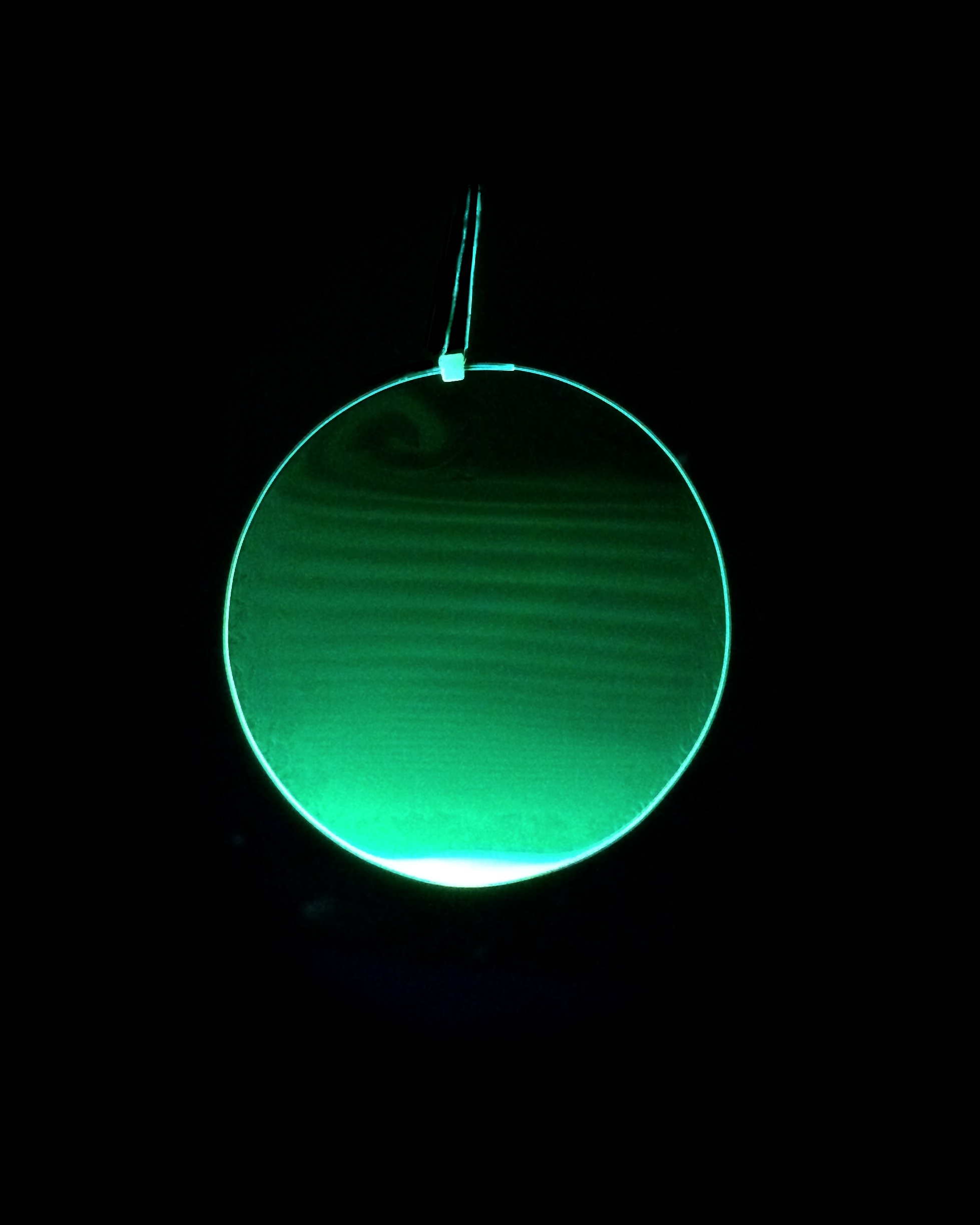}
\hspace{12pt}
\includegraphics[width=0.2125\textwidth]{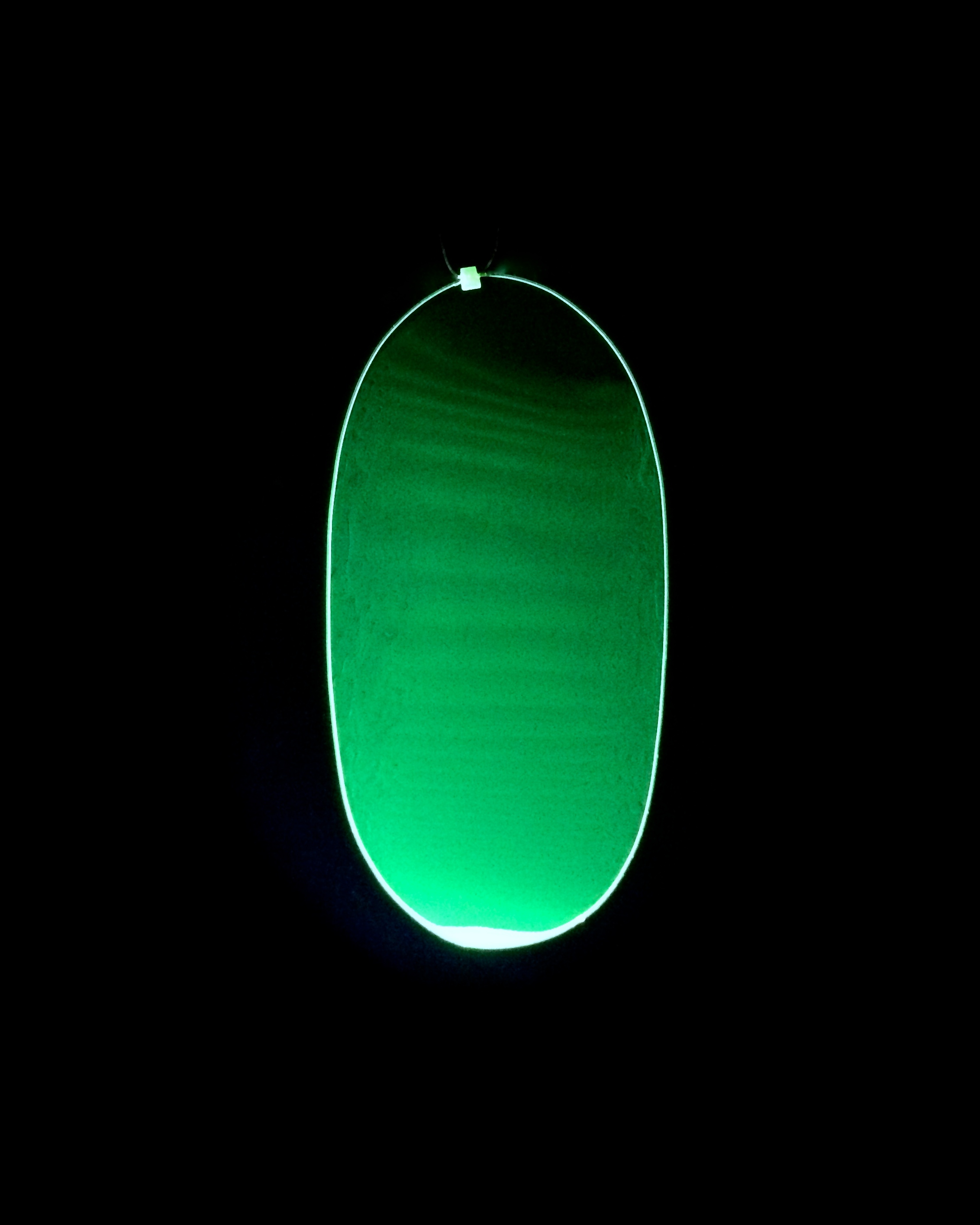}
\hspace{12pt}
\includegraphics[width=0.2125\textwidth]{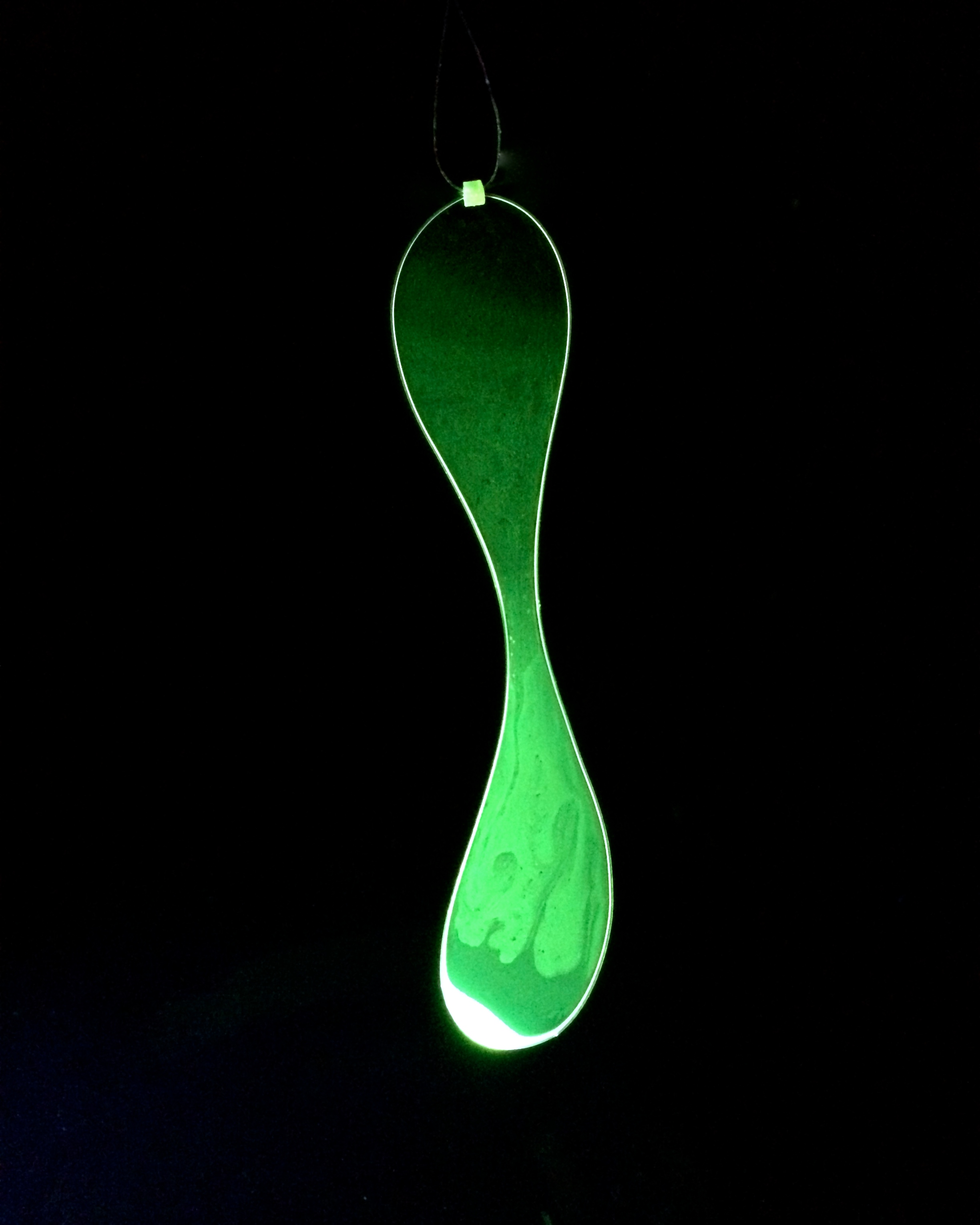}
\hspace{12pt}
\includegraphics[width=0.2125\textwidth]{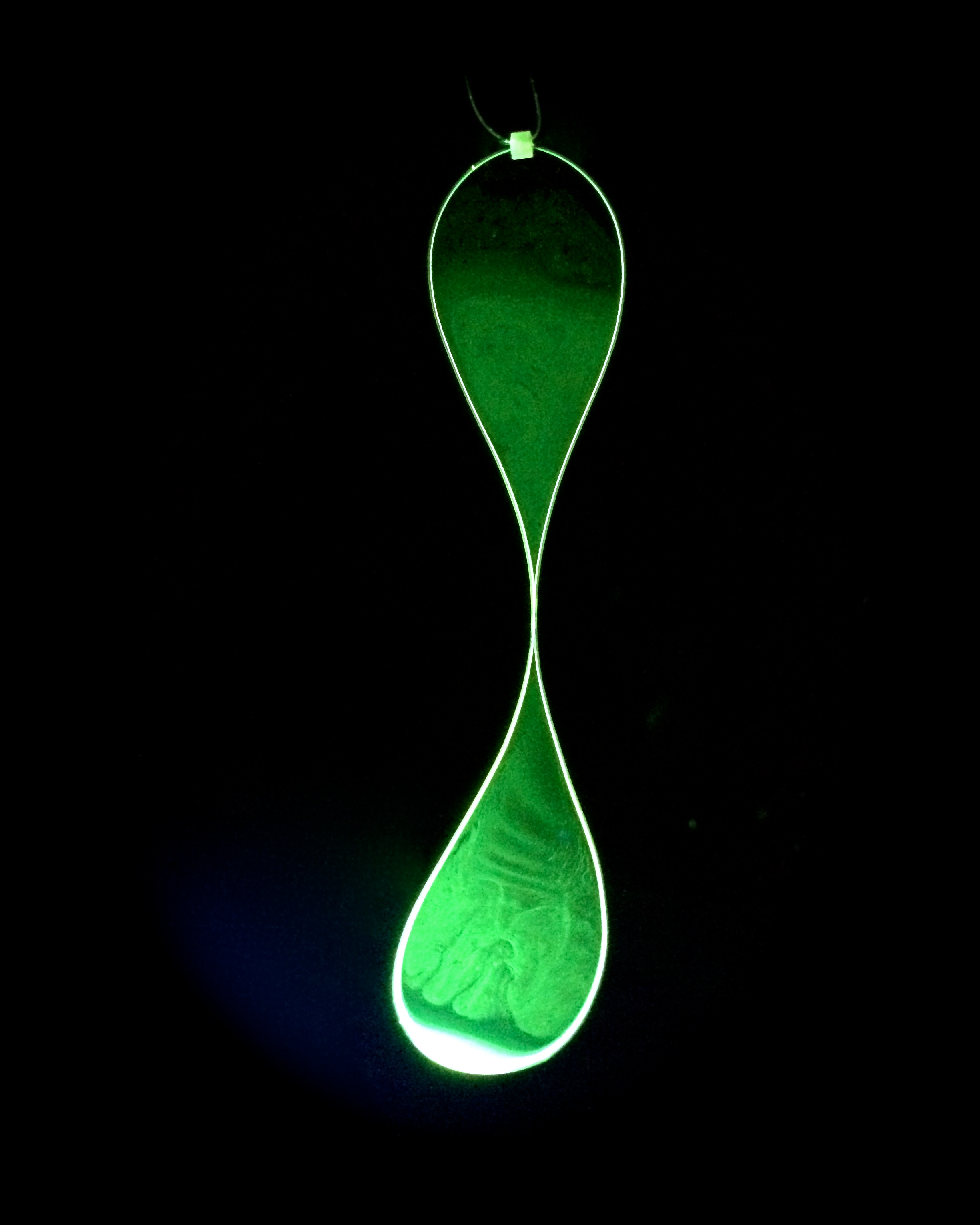}
\put(-450,110){\color{white}\small(a)}
\put(-332,110){\color{white}\small(b)}
\put(-212,110){\color{white}\small(c)}
\put(-94,110){\color{white}\small(d)}}
\\[12pt]
{\includegraphics[width=0.2125\textwidth]{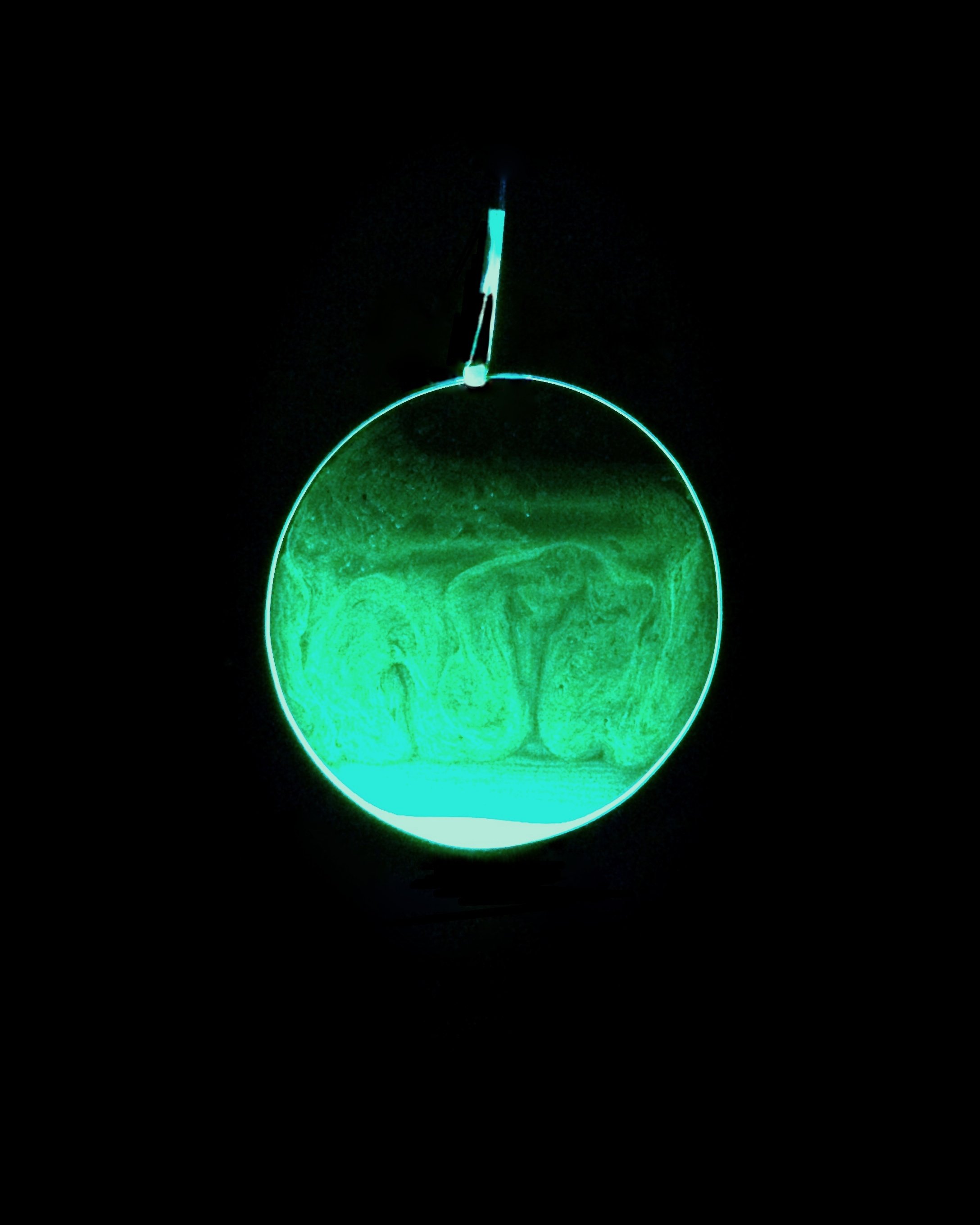}
\hspace{12pt}
\includegraphics[width=0.2125\textwidth]{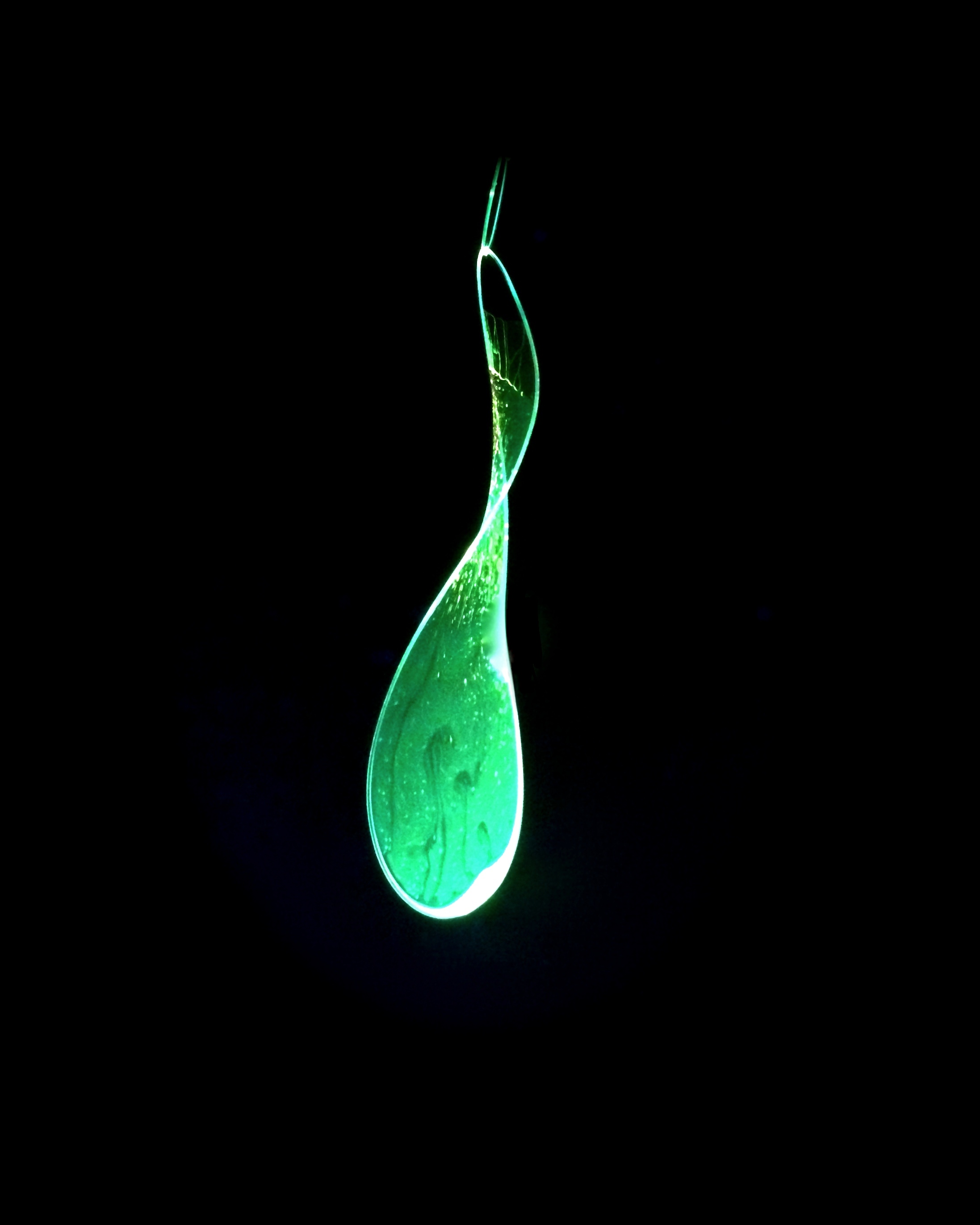}
\hspace{12pt}
\includegraphics[width=0.2125\textwidth]{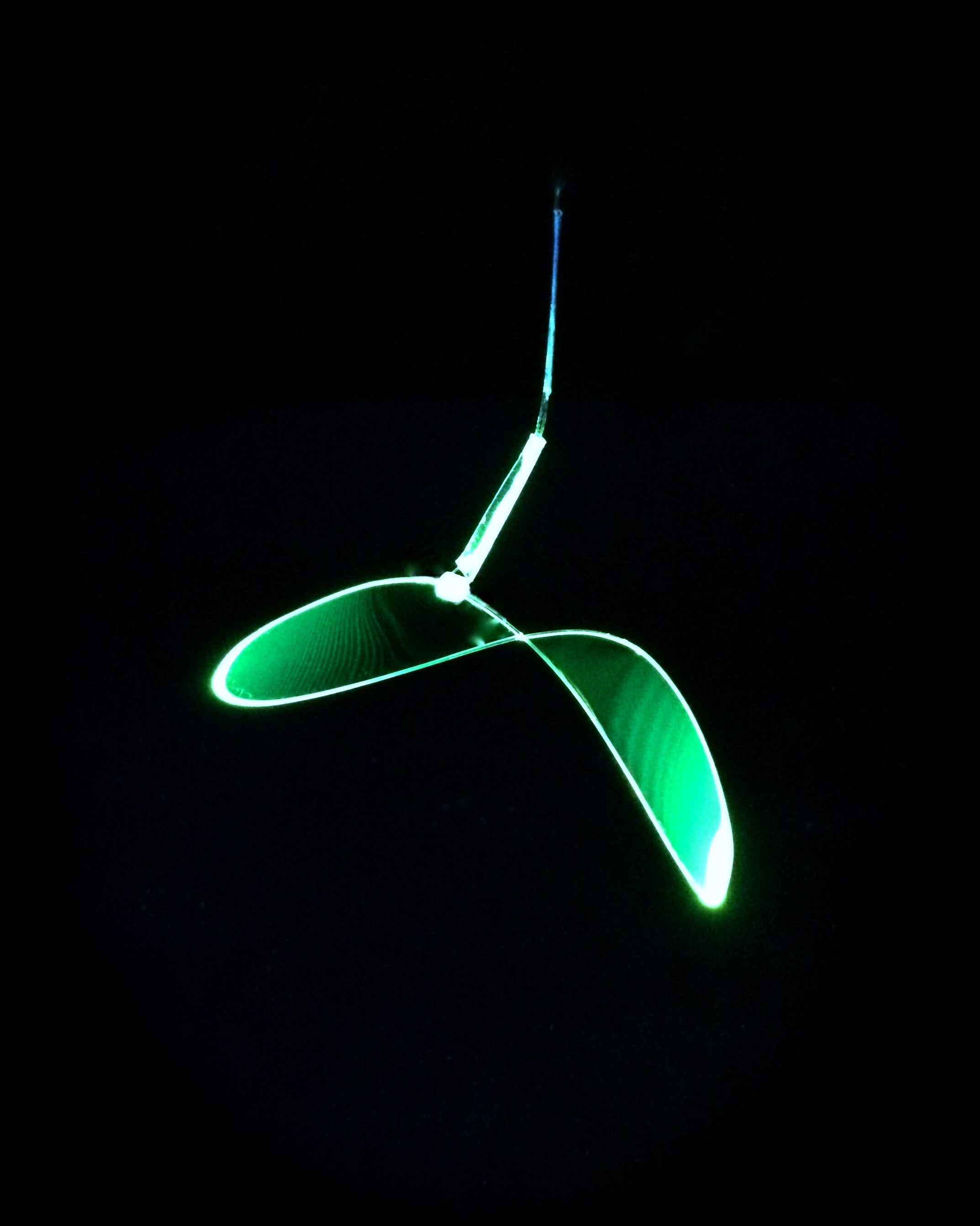}
\hspace{12pt}
\includegraphics[width=0.2125\textwidth]{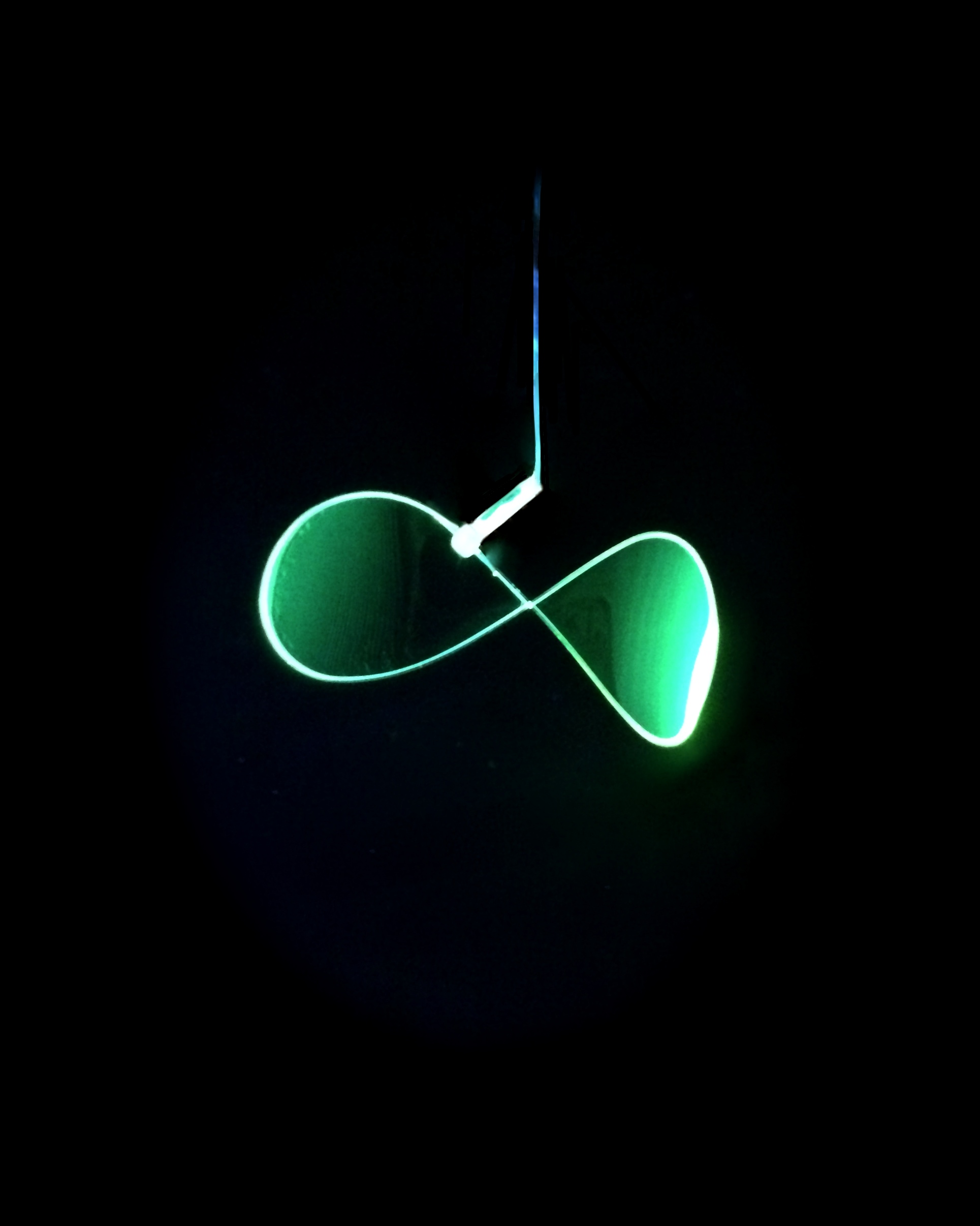}
\put(-450,110){\color{white}\small(e)}
\put(-332,110){\color{white}\small(f)}
\put(-212,110){\color{white}\small(g)}
\put(-94,110){\color{white}\small(h)}}
\caption{Various equilibrium configurations: In the first row, the loop are free of pre-twist and their lengths increase from left to right consistent with: (a) 17~cm; (b) 20~cm; (c) 21~cm; and (d) 22.5~cm. In the second row, the length of the fishing line is held constant at 17~cm and the twist density increases from left to right consistent with: (e) 0; (f) $\pi$; (g) $2\pi$; and (h) $4\pi$.}
\label{f4}
\end{figure}

Figure~\ref{f4} shows configurations for representative choices of the salient dimensionless parameters. Loops that are untwisted and sufficiently small adopt flat circular configurations (Figure~\ref{f4}(a)). Increasing the length of the loop slightly above a certain critical length results in flat ellipse-like configurations (Figure~\ref{f4}(b)). Further increasing the length of the loop leads to elongation and necking (Figure~\ref{f4}(c)), resulting eventually in a dumbell-like configuration in which the bounding loop exhibits a single point of self-contact. Still further increasing the length of the loop results in the extension of the point of self-contact to a line of self-contact (Figure~\ref{f4}(d)).

Untwisted loops which are small enough to be stable with respect to in-plane perturbations can be destabilized by imparting twist. The loops in Figures~\ref{f4}(e)--(h) all have the same length as the planar circle depicted in Figure~\ref{f4}(a). Nevertheless, twisting the bounding loop renders it unstable and is manifested by buckling to a saddle-shaped configuration (Figure~\ref{f4}(f)). With further twisting of the bounding loop, the saddle becomes increasingly non-planar and finally collapses to a twisted figure-eight (Figures~\ref{f4}(g) and (h)). 

The critical length at which an untwisted loop buckles from a circular configuration to an ellipse-like configuration was found to decrease as the cross-sectional diameter of the line was reduced. This was also the case for the critical length at which a point of self-contact forms. However, the buckling due to twisting depicted in Figure~\ref{f4}(e)--(h) was not observed to be sensitive to the cross-sectional diameter of the bounding loop. 

To provide additional insight, we consider the dimensionless parameters $\eta$, $\chi$, and $\mu$ defined in \eqref{dgroups1}. For a circular rod with cross-sectional diameter $d$, the bending rigidity is given by 
\be
a=IY=\frac{\pi d^4Y}{64},
\label{bending}
\ee
with $I$ the second area moment and $Y$ the Young's modulus, while the torsional rigidity is given by 
\be
c=JG=\frac{\pi d^4G}{34},
\ee
with $J$ the polar moment of inertia and $G$ the shear modulus. Whereas a change in the cross-sectional diameter of the loop changes the dimensionless parameter $\eta$ through the bending rigidity $a$, neither $\chi$ nor $\mu$ depends on $d${\clb ,} since $a$ and $c$ both scale with $d^4$. The ratio $\chi=c/a$ of the torsional rigidity to the bending rigidity does, however, depend on the Poisson's ratio $\nu$ of the selected material through
\be
\chi=\frac{1}{1+\nu}.
\ee
To estimate $\chi$ for the loops used in our experiments, the total twisting angle at which a loop was observed to buckle out of plane in the absence of a spanning soap film was measured for loops with different cross-sectional diameters, resulting in a mean critical total twisting angle $\Psi=(4.7 \pm 0.1) \pi$. Using \citeapos{Michell} formula $\Psi=2\pi R\Omega= 2\pi \sqrt{3} a/c$ for the critical twist density yields $\chi=0.737 \pm 0.015$. This value corresponds to a Poisson's ratio of $0.35 \pm 0.02$, which is in good agreement with the ratio reported for PVDF polymer by \citet{PVDF2}.

The bending rigidity of the PVDF fishing line used in our experiments can be calculated from substituting the cross-sectional diameter $d$ and measured value of the Young's modulus $Y$ in \eqref{bending}, which gives
\be
a=(3.09\pm0.11)\times10^{-7}~\text{Nm}^2. 
\label{aest}
\ee
However, the measurement of the Young's modulus and the relation \eqref{bending} for the bending rigidity both rest on assuming that the bounding loop is uniform and of circular cross-section. Motivated by this observation, we performed an alternative direct measurement of the bending rigidity to determine whether these assumptions are reasonable. Specifically, tests with untwisted loops of a given cross-sectional diameter were conducted and the critical length at which a point of self-contact forms was determined. The procedure was repeated with loops of two different cross-sectional diameters. With the measured values of surface tension $\sigma$ and Young's modulus $Y$, the critical lengths so determined were found to correspond to a mean value of the dimensionless parameter $\eta$ defined in \eqref{dgroups1}$_3$ equal to
\be
\eta_{\text{contact}}=3.09 \pm 0.18.
\label{nucontact0}
\ee
The distinction between the value of $\eta_{\text{contact}}$ in \eqref{nucontact0} and the value $\eta=3$ at which \citet{giomi2012} and \citet{chen2014} predicted that buckling from a flat circular configuration to a planar ellipse should occur seems to be unreasonably small and therefore raises doubts about the the utility of the value \eqref{aest} of $a$ obtained on the basis of assuming that the bounding loop is uniform and of circular cross-section. 

\citet{djondjorov2011} determined the critical hydrostatic pressure that enforces self-contact in elastic rings and tubes. On neglecting the slight non-planarity of the configuration of the loop and the spanning surface at the onset of self-contact and considering the analogy between their problem and the planar version of the current problem evident from \eqref{planar}, their results can be used to advantage here. Based on their solution,  self-contact in planar configurations of the current problem is expected at a critical value
\be
\eta_{\text{contact}}=5.247,
\label{nucontact}
\ee
which differs significantly from the value \eqref{nucontact0} obtained previously. This suggests that the previously measured bending rigidity, which resulted from substituting the Young's modulus obtained from a  tensile test into \eqref{bending}, should be reduced by a multiplicative correction factor of 0.59, giving
\be
a_{\text{corrected}}=(1.96 \pm 0.06) \times 10^{-7}\text{~Nm}^2.
%E_{\text{corrected}}=1.35\pm0.02\text{~GPa}.
\ee

Finally, the measured parameters for surface tension $\sigma$ and the bending and twisting rigidities $a$ and $c$, together with the correction described above, were used to determine ordered pairs $(\mu,\eta)$ of the dimensionless parameters associated with the onset of instability for twisted elastic loops spanned by soap films. The result is depicted in Figure~\ref{f3}, which also includes a plot of the theoretical prediction. The diagram obtained from the measured parameters follows a qualitative trend consistent with that suggested by theory. On applying the correction factor of $0.59$ leading to \eqref{nucontact}, good quantitative agreement is also achieved.
{   
\begin{figure}[!t]
\centering
\includegraphics [width=10cm]{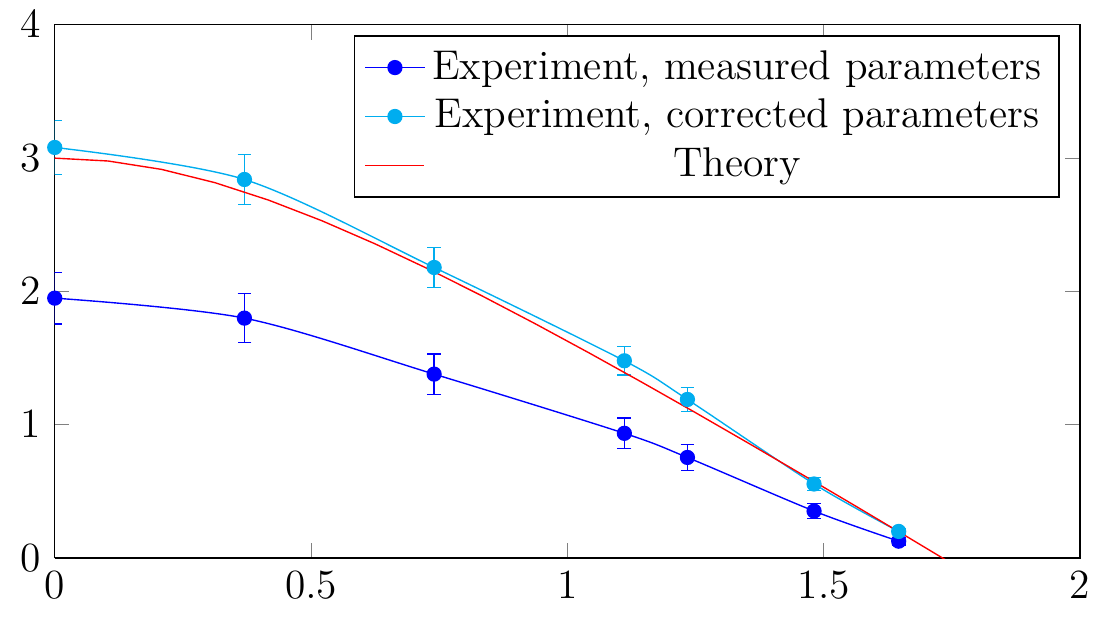}
\put(-300,80){$\eta$}
\put(-144,-8){$\mu$}
%\begin{tikzpicture}
%    \begin{axis}[
%    domain=0:2.5,
%     xmin=0, xmax=2,
%    ymin=0, ymax=4,
%        xlabel=$\mu$,
%        ylabel=$\nu$,
%        width=12cm, height=7cm,
%         ytick={0,1,2,3,4},
%        xtick={0,0.5,1,1.5,2},]
%  \addplot[
%    color=blue,
%   smooth,
%    mark=*,
%    /pgfplots/error bars/.cd,
%        x dir=both,
%        y dir=both,
%        y explicit
%] table [
%    x=xValue1,
%    y=yValue1,
%    y error=Deltay1,
%] {Table.txt};
% \addlegendentry{Experiment, measured parameters}
%
%
%\addplot [
%    color=cyan,
%   smooth,
%    mark=*,
%    /pgfplots/error bars/.cd,
%        x dir=none,
%        y dir=both,
%        y explicit
%] table [
%    x=xValue2,
%    y=yValue2,
%    y error=Deltay2
%] {Table.txt};
% \addlegendentry{Experiment, corrected parameters}
% %
%    \addplot[color=red]{4.5-1.5*sqrt(1+8/3*x^2)};
%    \addlegendentry{Theory}
%  
%    \end{axis}
%    \end{tikzpicture}
\caption{The pairs of dimensionless parameters $\mu$ and $\eta$ which render a planar circular configuration unstable, obtained with the experimentally measured bending rigidity $a=(3.09 \pm 0.11) \times 10^{-7}$~Nm$^2$ ({\color{blue}$\bullet$}) and with the corrected value $a_{\text{corrected}}=(1.96 \pm 0.06) \times 10^{-7}$~Nm$^2$ ({\color{cyan}$\bullet$}).
%     length parameter $\sigma/E=(1.00 \pm 0.07)\times10^{-11}$~m ({\color{blue}$\bullet$}) and with the corrected length parameter $\sigma/E=(1.56 \pm 0.08)\times10^{-11}$~m ({\color{cyan}$\bullet$}). 
     While the blue and cyan curves indicate polynomial fits to the experimental data, the red curve represents the theoretical prediction.}
\label{f3}
\end{figure}
%

%%%%%%%%%%%%%%%%%%%%%%%%%%%%%%%%%
\section{Conclusions}
\label{Conclusion}
The stability of soap films spanning closed elastic rods of constant length was studied. The energy of the system consists of the lineal bending and twisting energies of the bounding loop and the areal energy of the soap film. The first energy variation vanishes at equilibrium, resulting in {\clr the} Euler--Lagrange equations for the surface {\clr and} its boundary. The stability of a given equilibrium configuration is an indicator of whether or not that configuration is indeed observable. Motivated by this, the second-variation condition was used to derive  stability conditions. Three dimensionless parameters enter the stability conditions. Of these, the parameter $\eta$ familiar from the Euler--Plateau problem measures the importance of the areal energy relative to the lineal bending energy. This parameter depends on the surface tension of the film and the length and the bending rigidity of the bounding loop. The parameter $\chi$ reflects the importance of the torsional rigidity of the loop relative to its bending rigidity. The parameter $\mu$ subsumes information concerning $\chi$ and twist density of the loop. Our stability conditions were applied to examine the response of a flat circular ground state with respect to planar and transverse perturbations, in which case $\chi$ influences stability only in combination with the dimensionless twist density through $\mu$. Since the influences of radial and transverse perturbations are coupled through $\mu$, the two separate stability conditions obtained for the Euler--Plateau problem by \citet{chen2014} are incorporated here in a single condition. As might be expected, for a loop of given length and cross-sectional diameter spanned by a soap film, the critical twist density is bounded above by \citeapos{Michell} critical value for twisted ring. Increasing either of the total twist of the bounding loop, the size of the loop, or the surface tension of the film destabilizes a circular disk. 

Experiments were performed with fishing lines dipped into and removed from soapy water with the dual objective of gaining insight on configurations that arise from both small and large deformations and on the stability of various equilibrium configurations. In the first group of experiments, no twist was imparted to the loops and the effect of increasing the parameter $\eta$ by increasing the length of the bounding loop was studied. Loops that are untwisted and sufficiently short adopt circular configurations. On increasing the length of the loop and, hence, $\eta$, these change into ellipse-like configurations. Further increasing the length of the loop leads to elongation and necking of such configurations, resulting eventually {\clr in} a point of self contact. This point of self-contact expands to a line if the parameter $\eta$ is increased even further. In a second group of experiments, the parameter $\eta$ was held fixed and the influence of changing the parameter $\mu$ was considered. For $\mu\neq0$, the buckling analysis of \citet{biria2014} predicts two families of saddle-like modes that bifurcate from a circular disk, the difference being that one family projects onto a circular domain and the other projects onto an elliptical domain. Only the latter modes were observed in the experiments, which suggests that the former modes are unstable. On imparting larger twist densities to the bounding loop, a saddle buckles to a twisted figure-eight configuration. 

To conduct a quantitative comparison between theory and experiment, various material and geometrical parameters determining $\eta$ and $\mu$ were measured. The measurement of the Young's modulus and, consequently, of the bending rigidity, was predicated on the restriction to bounding loops with uniform circular cross sections. The first group of experiments demonstrated a discrepancy between the critical value of the parameter $\eta$ at which self-contact develops and theoretical predictions. Considering that the length of the loop and the surface tension of the soap film were measured directly and with high precision, that discrepancy raises concerns about the accuracy of the measurement of the bending rigidity, and hence, the grounds for assuming that the loops used in our experiments were uniform with circular cross-sections. With this in mind, the results of a first group of experiments were used to obtain a corrective scaling to the bending rigidity. The measured parameters and the corrective scaling were employed in a final group of experiments, with the objective of constructing an associated stability diagram. The analytical predictions and experimental results are in good qualitative agreement. With the corrective scaling, the results agree quantitatively as well.

The problem of a twisted elastic loop spanned by a soap film involves an interplay between surface and edge energies reminiscent of that present in various biological systems. In particular, the problem is closely related to the behavior of lipid membranes bounded by chiral protein chains, such as discoidal high-density lipoprotein (HDL) particles. Therefore, the equilibrium shapes observed in the experiments are expected to have similar biological counterparts for relevant ranges of the two dimensionless bifurcation parameters. Indeed, \citet{catte2006} have reported analogs of the observed saddle and the figure-eight conformations observed here for discoidal HDL particles.

%%%%%%%%%%%%%%%%%%%%%%%%%%%%%%%%
%\subsection*{Acknowledgement}
%\label{Acknowledgement}
%We are grateful to Professors Amy Shen and Mahesh Bandi, for providing access to their laboratory instruments and for their many fruitful suggestions. Help from other members of the Micro/Bio/Nanofluidics and the Collective Interactions units at Okinawa Institute of Science and Technology is also gratefully appreciated. We are also thankful to the Fiber Physics and Quality Control Laboratory at Amirkabir University of Technology for granting us access to their instruments.
%%%%%%%%%%%%%%%%%%%%%%%%%%%%%%%%%

\appendix

\section{Detailed derivations of the first and second variations}

Consider an infinitesimal virtual displacement of $\calS$ to $\calS_\eps$, with $\eps>0$ being a small, dimensionless parameter. Let $\varrho_\eps$ denote the corresponding value of a quantity $\varrho$ defined on $\calS$ or its boundary $\calC$. The variation $\delta\varrho$ of $\varrho$ is then given by
\begin{equation}
\delta{\varrho}=\lim_{\eps\to0}\frac{\varrho_\eps-\varrho}{\eps}.
\label{var}
\end{equation}
To calculate the variations of quantities defined on $\calS$, which we refer to as areal quantities, and quantities defined on $\calC$, which we refer to as lineal quantities, we follow techniques established by \citet{virga1995variational},  \citet{rv99}, and  \citet{PhysRevE.73.021602}.

\subsection{Preliminary identities}
Assume that $\calS$ is a material surface embedded in an arbitrarily chosen three-dimensional body $\calP$. Consider a virtual deformation $\bfchi_\eps$ that alters $\calP$ to $\calP_\eps$ by mapping a generic point in $\calP$ with position vector $\bfr$ to a point in $\calP_\eps$ with position vector
\be
\bfr_\eps=\bfchi_\eps(\bfr)
=\bfr+\eps\bfv+\half\eps^2\tilde\bfv+o(\eps^2){\color{red},}
\label{ap1}
\ee
where $\bfv$ and $\tilde\bfv$ denote the first and second variations of $\bfr$ and are of order unity. The virtual deformation gradient
\be
\bfF_\eps=\nabla\bfchi_\eps(\bfx)
=\idem+\eps\nabla{\bfv} +\half \eps^2 \nabla{\tilde\bfv} + o(\eps^2),
\label{ap2}
\ee
determines the virtual areal and lineal stretches, provided in the next sections, and thus is of considerable importance.

Recall that $\bfm$ denotes the unit normal field to $\calS$ and, thus, that the perpendicular projector to $\calS$ is given by 
\be
%\bfP=
\idem-\bfm\otimes\bfm,
\label{id1}
\ee
where $\idem$ is the three-dimensional identity tensor on $\calP$. Suppose that $\bfw$ is a smooth field defined on $\calP$ and let $\bsw$ denote the restriction of $\bfw$ to $\calS$. The surface gradient $\grads\bfw$ of the restriction of $\bfw$ to $\calS$ is given in terms of the restriction of its three-dimensional gradient $\nabla\bfw$ and $\bfm$ by 
\be
\grads\bfw=(\nabla\bfw)(\idem-\bfm\otimes\bfm).
\label{id2}
\ee
Similarly, the surface divergence $\divs\bfw$ of the restriction of $\bfw$ to $\calS$ is given by
\be
\divs\bfw=\tr(\grads\bfw)
=\Div\bfw-\bfm\cdot(\nabla\bfw)^{\trans}\bfm
=\Div\bfw-\bfm\cdot(\nabla\bfw)\bfm.
\label{id3}
\ee
Useful consequences of \eqref{id1} and \eqref{id2} are 
\begin{align}
\tr[(\grads\bfw)^2]&=\tr[(\nabla\bfw-(\nabla\bfw)\bfm\otimes\bfm)^2]
\gogo
&=\tr[(\nabla\bfw)^2]-2\bfm\cdot(\nabla\bfw)^2\bfm+(\bfm\cdot(\nabla\bfw)\bfm)^2,
\label{id4}
\end{align}
and
\begin{align}
(\tr(\grads\bfw))^2&=[\tr(\nabla\bfw-(\bfm\otimes\bfm)(\nabla\bfw)^{\trans}]^2
\gogo
&=[\tr(\nabla\bfw-\bfm\otimes(\nabla\bfw)\bfm)]^2
%\gogo
%&=[\tr(\nabla\bfw)-\bfm\cdot(\nabla\bfw)\bfm]^2
\gogo
&=[\tr(\nabla\bfw)]^2-2(\Div\bfw)\bfm\cdot(\nabla\bfw)\bfm+(\bfm\cdot(\nabla\bfw)\bfm)^2,
\label{id5}
\end{align}
along with 
\begin{align}
\abs{(\grads\bfw)^{\trans}\bfm}^2
&=\abs{(\nabla\bfw)^{\trans}\bfm-(\bfm\cdot(\nabla\bfw)\bfm)\bfm}^2
\gogo
%&=\abs{(\nabla \bsw)^{\trans} \bfm}^2-2[(\nabla \bsw)^{\trans}\bfm]\cdot(\bfm \otimes \bfm (\nabla \bsw)^{\trans}\bfm)+\abs{\bfm \cdot (\nabla \bsw)^{\trans} \bfm}^2
%\gogo
&=\abs{(\nabla\bfw)^{\trans}\bfm}^2-(\bfm\cdot(\nabla\bfw)\bfm)^2.
\label{id6}
\end{align}

As a particular consequence of \eqref{id4} and \eqref{id5}, the second invariant $I_2(\grads\bfv)$ of the surface gradient $\grads\bfv$ of the restriction of $\bfv$ to $\calS$ is given in terms of the restriction of its three-dimensional gradient $\nabla\bfv$ to $\calS$ and $\bfm$ by 
\be
I_2(\grads\bfv)=I_2(\nabla\bfv)+\bfm\cdot(\nabla\bfv)^2\bfm
-(\Div\bfv)\bfm\cdot(\nabla\bfv)\bfm.
\label{id7}
\ee

%%%%%%%%%%%%%%%%%%%%%%%%%%%%%
\subsection{Areal quantities}

%The purpose of this section is to derive the first and second variation of the surface integral appearing in \eqref{EL1}. First some identities that relate 

Consider the virtual areal Jacobian
\be
\jmath_\eps=\abs{\bfF^c_\eps \bfm},
%=(\det\bfF_\eps)|\bfF_\eps^{\rm -\trans}\bfm|,
\label{ajac}
\ee
of $\calS_\eps$, where, given an arbitrary tensor $\bfA$,
\be
\bfA^c=(\bfS^2-(\tr\bfS)\bfS+I_2(\bfS)\idem)^{\trans},
\label{iden2}
\ee
with
\be
I_2(\bfS)=\half[(\text{tr}\bfS)^2-\text{tr}(\bfS^2)],
\label{iden2}
\ee
denotes its cofactor. Then, since $(\idem+\bfS)^c=(1+\tr\bfS)\idem-\bfS^{\trans}+\bfS^c$,  
\begin{multline}
\bfF^c_\eps =\idem+\eps[(\Div\bfv)\idem-(\nabla\bfv)^{\trans}\mskip1.5mu] 
+\eps^2[\half(\Div\tilde\bfv)\idem-\half(\nabla\tilde\bfv)^{\trans}
\\[4pt]+I_2(\nabla\bfv)\idem
+(\nabla\bfv)^2-(\Div\bfv)\nabla\bfv]+o(\eps^2),
\label{ap6}
\end{multline}
from which it follows that
\begin{multline}
\abs{\bfF^c_\eps \bfm}=\idem+\eps(\Div\bfv-\bfm\cdot(\nabla\bfv)\bfm)
\\[4pt]
+\half\eps^2[\Div\tilde\bfv-\bfm\cdot\nabla\tilde\bfv^{\trans}\bfm
+\abs{(\nabla\bfv^{\trans})\bfm}^2-(\bfm\cdot(\nabla\bfv)\bfm)^2
\\[4pt]+2(I_2(\nabla\bfv)+\bfm\cdot(\nabla\bfv)^2\bfm 
 -(\Div\bfv)\bfm\cdot(\nabla\bfv)\bfm)
]+o(\eps^2).
\label{ap7}
\end{multline}
Next, using \eqref{id3}, \eqref{id6}, and \eqref{id7} in \eqref{ap7} results in
\be
\jmath_\eps=\abs{\bfF^c_\eps \bfm}=1+\eps{\divs{\bfv}}+\half \eps^2[\divs{\tilde\bfv}+ \abs{(\grads \bfv)^{\trans}\bfm}^2+2I_2(\grads \bfv)]+o(\eps^2).
\label{ap8}
\ee

Since $\int_{\calS_\eps}\sigma=\int_{\calS}\sigma\jmath_\eps$, it follows from \eqref{ap8} that
\be
\int_{\calS_\eps} \sigma=\int_{\calS}\sigma+\eps\int_{\calS}\sigma{\divs{\bfv}}+\eps^2\int_{\calS}\sigma(\divs{\tilde\bfv}+\half \abs{(\grads \bfv)^{\trans}\bfm}^2+I_2(\grads \bfv)),
\label{ap9}
\ee
which, on applying the definition \eqref{var} of the first variation and the surface divergence theorem, yields 
\be
\del \int_{\calS} \sigma=\int_{\calS}\sigma{\divs{\bfv}}
=\int_{\calC}\sigma{\bfv}\cdot\bfnu-\int_{\calS}2\sigma H \bfv \cdot \bfm
\label{ap10}
\ee
for the first variation of the surface energy integral. Similarly,
\be
\del^2 \int_{\calS} \sigma=\int_{\calS}\sigma[\abs{\grads \bfv}^2+2I_2(\grads \bfv)-4H \tilde\bfv \cdot \bfm]+ \int_{\calC} 2\sigma{\tilde\bfv} \cdot \bfnu. 
\label{ap11}
\ee
%

%%%%%%%%%%%%%%%%%%%%%%%%
\subsection{Lineal quantities}
\subsubsection{Consequences of inextensibility}
The virtual lineal stretch of $\calC$, denote by $\lambda_\eps$, is related to the virtual deformation gradient by 
\be
\lambda_\eps=|\bfF_\eps\bfe|,
\label{ap3}
\ee
which on using the expansion \eqref{ap2} and relation $(\nabla{\bfu})\bfe={\bfu}'$, results in
\be
\lambda_\eps=1+\eps \bfv' \cdot \bft +\half \eps^2[\abs{\bfv'}^2+ 2 {\tilde \bfv}' \cdot \bft -\abs{\bfv' \cdot \bft}^2]+o(\eps^3).
\label{ap4}
\ee
From \eqref{ap4}, imposing the local inextensibility on $\calC$ up to second order in $\eps$ requires that
\be
\del \lambda=\bfv' \cdot \bft =0
\qquad\text{and}\qquad
\del^2 \lambda= \abs{\bfv'}^2+ 2 {\tilde\bfv}' \cdot \bft -\abs{\bfv' \cdot \bft}^2=0.
\label{inex1}
\ee

Let $g$ be defined on $\calC$. Then, by \eqref{inex1}$_1$,
\be
\del(g')=(\del g)'+\del\lambda g'=(\del g)'.
\label{inex2}
\ee
Additionally, 
\be
\del\int_{\calC}g=\int_{\calC}(\del g+g\del\lambda)=\int_{\calC}\del g.
\label{inex3}
\ee
%

%%%%%%%%%%%%%%%%%%%%%%%%%%%%%%%%%%%%%%%%
\subsubsection{First variations}

%{\color{red}$\del\kappa$, $\del\bfn$, $\del\tau$, $\del\bfb$, $\del\psi$, $\del\Omega$}

An immediate consequence of \eqref{inex2} is that the first variation $\del\bft$ of the tangent $\bft$ to $\calC$ can be expressed as
\be
\del\bft=\del(\bfr')=(\del\bfr)'=\bfv'.
\label{vart}
\ee
By \eqref{SF}$_1$, \eqref{inex2}, and \eqref{vart},
\be
\bfv''=(\del\bft)'=\del(\bft')=\del(\kappa\bfn)=(\del\kappa)\bfn+\kappa\del\bfn.
\label{vprimeprime}
\ee
Since $\bfn$ is unit-vector valued, $\bfn\cdot\del\bfn=0$, dotting both sides of \eqref{vprimeprime} with $\bfn$ yields the first variation $\del\kappa$ of the curvature $\kappa$ of $\calC$ as
\be
\del\kappa=\bfn\cdot\bfv''.
\label{varkappa}
\ee
Thus, by \eqref{curvec} and \eqref{varkappa}, the product $\kappa\del\kappa$ of $\kappa$ and its first variation $\del\kappa$ can be expressed as
\be
\kappa\go\del\kappa=\bfk\cdot\bfv''
=(\bfk\cdot\bfv'-\bfk'\cdot\bfv)'+\bfk''\cdot\bfv.
\label{vark}
\ee
Further, solving for $\kappa\go\del\bfn$ in \eqref{vprimeprime} and using \eqref{varkappa} gives
\be
\kappa\go\del\bfn=\bfv''-(\bfn\cdot\bfv'')\bfn.
\label{varn}
\ee
%
%{\color{red}Determine a relation for $(\del\bfn)'$\dots}

Next, since $\bfb$ is unit-vector valued, $\bfb\cdot\del\bfb=0$ and
\be
\bfb\cdot\del(\tau\bfb)=\bfb\cdot(\tau\del\bfb+(\del\tau)\bfb)=\del\tau.
\label{vartau0}
\ee
Bearing in mind that $\bfb\cdot\bft=0$ and, since $\bfn$ is unit-vector valued, $\bfn\cdot\del\bfn=0$, \eqref{SF}$_2$, \eqref{SF}$_3$, \eqref{inex2}, and \eqref{vartau0} yield and expression for the first variation $\del\tau$ of the torsion $\tau$ of $\calC$ in the form
\begin{align}
\del\tau&=\bfb\cdot\del(\bfn'+\kappa\bft)
\notag\\[4pt]
&=\bfb\cdot(\del\bfn)'+(\del\kappa)\bfb\cdot\bft+\kappa\bfb\cdot\del\bft
\notag\\[4pt]
&=(\bfb\cdot\del\bfn)'-\bfb'\cdot\del\bfn+\kappa\bfb\cdot\del\bft
\notag\\[4pt]
&=(\bfb\cdot\del\bfn)'+\tau\bfn\cdot\del\bfn+\kappa\bfb\cdot\del\bft
=(\bfb\cdot\del\bfn)'+\kappa\bfb\cdot\del\bft.
\end{align}
Thus, by \eqref{vart} and \eqref{varn},
\begin{align}
\del\tau&=(\kappa^{-1}\bfb\cdot(\bfv''-(\bfn\cdot\bfv'')\bfn))'+\kappa\bfb\cdot\bfv'
\notag\\[4pt]
&=(\bfb\cdot(\kappa\bfv+\kappa^{-1}\bfv''))'-(\kappa\bfb)'\cdot\bfv.
\label{vartau}
\end{align}

Next, by \eqref{twist}, \eqref{vandiota}$_2$, \eqref{inex2}, and \eqref{vartau}, the first variation $\del\Omega$ of the twist density $\Omega$ of $\calC$ is given by
\begin{align}
\del\Omega&=\del\tau+\del(\psi')
\notag\\[4pt]
&=\del\tau+(\del\psi)'=[\bfb\cdot(\kappa\bfv+\kappa^{-1}\bfv'')]'-(\kappa\bfb)'\cdot\bfv+\iota',
\label{varomega1}
\end{align}
from which it follows that the product $\Omega\mskip1.5mu\del\Omega$ can be expressed as 
\begin{multline}
\Omega\mskip1.5mu\del\Omega
%&=(\Omega(\mskip1.5mu\del \psi+\kappa^{-1} \bfb \cdot \bfv''+\kappa \go \bfb \cdot \bfv))'-\Omega' (\del \psi+\kappa^{-1} \bfb \cdot \bfv'') -(\Omega \mskip1.5mu \kappa \go \bfb)' \cdot \bfv \gogo
=(\Omega(\bfb\cdot(\kappa\bfv+\kappa^{-1}\bfv'')+\iota)+\kappa^{-1}\Omega'\bfb\cdot\bfv'
-(\kappa^{-1}\Omega'\bfb)'\cdot\bfv)'
\\[4pt]
+((\kappa^{-1}\Omega' \bfb)'-\kappa\Omega\bfb))' \cdot\bfv-\Omega'\iota.
\label{varomega2}
\end{multline}

Since $\bfb$ is unit-vector valued, $\bfb\cdot\del\bfb=0$ and $\del\bfb$ must obey
\begin{align}
\del\bfb&=(\idem-\bfb\otimes\bfb)\del\bfb
\notag\\[4pt]
&=(\bft\otimes\bft+\bfn\otimes\bfn)\del\bfb
\notag\\[4pt]
&=(\bft\cdot\del\bfb)\bft+(\bfn\cdot\del\bfb)\bfn.
\label{varb0}
\end{align}
Further, since $\bfb\cdot\bft=0$ and $\bfb\cdot\bfn=0$, \eqref{vart} and \eqref{varn} yield
\begin{align}
\bft\cdot\del\bfb&=-\bfb\cdot\del\bft
\notag\\[4pt]
&=-\bfb\cdot\bfv'
\label{varbt}
\end{align}
and
\begin{align}
\bfn\cdot\del\bfb&=-\bfb\cdot\del\bfn
\notag\\[4pt]
&=-\kappa^{-1}\bfb\cdot(\bfv''-(\bfn\cdot\bfv'')\bfn)
\notag\\[4pt]
&=-\kappa^{-1}\bfb\cdot\bfv'',
\label{varbn}
\end{align}
from which it follows that
\be
\del\bfb=-(\bfb\cdot\bfv')\bft-\kappa^{-1}(\bfb\cdot\bfv'')\bfn.
\label{varb}
\ee

Finally, consider the first variation $\del(\kappa\bfb)$ of the product of the curvature $\kappa$ and the binormal $\bfb$ of $\calC$. Since $\del(\kappa\bfb)=(\del\kappa)\bfb+\kappa\del\bfb$, \eqref{varkappa} and \eqref{varb} give
\begin{align}
\del(\kappa\bfb)
&=(\bfn\cdot\bfv'')\bfb-(\bfb\cdot\bfv')\kappa\bft-(\bfb\cdot\bfv'')\bfn
\notag\\[4pt]
&=(\bfb\otimes\bfn-\bfn\otimes\bfb)\bfv''-(\bfb\cdot\bfv')\kappa\bft,
\label{delkappab0}
\end{align}
which, since $\bfb\otimes\bfn-\bfn\otimes\bfb=(\bfn\times\bfb)\times=\bft\times$, results in
\be
\del(\kappa\bfb)%=(\bfn\times\bfb)\times\bfv''-(\bfb\cdot\bfv')\kappa\bft
=\bft\times\bfv''-(\bfb\cdot\bfv')\kappa\bft.
\label{delkappab}
\ee

%%%%%%%%%%%%%%%%
\subsubsection{Second variations}

Since $\del^2(\half\kappa^2)=(\del\kappa)^2+\kappa\del^2\kappa$, \eqref{vark} and \eqref{varn} yield
%The second variation of the contribution of bending to the lineal energy follows from \eqref{vark} and \eqref{varn} and one step of integration by parts:
 %
\be
\del^2(\half\kappa^2)
=(\bfn \cdot \bfv'')^2+\kappa(\del\bfn\cdot\bfv''+\bfn\cdot\del\bfv'')
=\abs{ \bfv''}^2+(\bfk\cdot\tilde\bfv'-\bfk'\cdot\tilde\bfv)'+\bfk''\cdot\tilde\bfv,
\label{var2k1}
\ee
which, in view of the periodicity of $\bfk$ and $\tilde\bfv$, leads to 
\be
\del^2 \int_{\calC} \half a \kappa^2= \int_{\calC}( \abs{ \bfv''}^2+\bfk''\cdot \tilde\bfv).
\label{var2k2}
\ee

Next, consider the second variation of twist energy density $ \del^2 (\half \Omega^2)=(\del\Omega)^2+\Omega \del^2 \Omega$ and invoke \eqref{twist} to yield
\begin{align}
 \del^2  (\half \Omega^2) &=(\del\Omega)^2+\Omega(\del^2\tau+\del \iota') \gogo
 &=(\del\Omega)^2+\Omega\del^2\tau+(\Omega \del\iota)'-\Omega' \del \iota.
 \label{var2omega1}
   \end{align}
Notice that, by \eqref{vartau},
\be
 \del^2\tau=(\del(\kappa^{-1}\bfb\cdot\bfv''))'+(\del(\kappa\bfb))\cdot\bfv'+\kappa\bfb\cdot\tilde{\bfv}',
   \ee
which, on substituting from \eqref{delkappab} and making use of \eqref{inex1}$_1$, yields
\be
 \del^2\tau=(\del(\kappa^{-1}\bfb\cdot\bfv'')+\kappa\bfb\cdot\tilde{\bfv})'+(\bft\times\bfv'')\cdot\bfv'-(\kappa\bfb)'\cdot\tilde{\bfv}.
 \label{var2tau}
\ee
Using \eqref{varomega1} and \eqref{var2tau} in \eqref{var2omega1} results in
\begin{multline}
 \del^2  (\half \Omega^2) =((\kappa^{-1} \bfb \cdot \bfv'')'+\kappa \go \bfb \cdot \bfv'+\iota')^2 
 +\Omega ((\bfv'' \times \bfv') \cdot \bft-(\kappa \go \bfb)' \cdot \tilde \bfv)
 \\[4pt]
 +[\Omega(\del(\kappa^{-1}\bfb\cdot\bfv'')+\kappa\bfb\cdot\tilde{\bfv}+\del \iota)]'-\Omega'[\del(\kappa^{-1}\bfb\cdot\bfv'')+\kappa\bfb\cdot\tilde{\bfv}+\del \iota],
 \label{var2omega2}
   \end{multline}
where, in view of the periodicity of $\iota$, $\kappa$, $\bfb$, $\bfv$ and $\tilde\bfv$ on $\calC$, an expression for the second variation of total contribution of the twist energy follows in the form
\begin{multline}
\del^2 \int_{\calC} \half c \Omega^2 =\int_{\calC} ((\kappa^{-1} \bfb \cdot \bfv'')'+\kappa \go \bfb \cdot \bfv'+\iota')^2 
 +\Omega ((\bfv'' \times \bfv') \cdot \bft-(\kappa \go \bfb)' \cdot \tilde \bfv)
 \\[4pt]
 -\Omega'[\del(\kappa^{-1}\bfb\cdot\bfv'')+\kappa\bfb\cdot\tilde{\bfv}+\del \iota].
 \label{var2omega2}
\end{multline}
Bearing in mind that 
\begin{align}
 (\del \Omega)^2& = 2 \iota' \del \tau+(\del \tau)^2+\iota'^2 
 \gogo&=2 \iota' (\del \Omega-\iota')+(\del \tau)^2+(\iota')^2 \gogo
  &=(2\iota \del \Omega)'-2 \iota \del \Omega'+(\del \tau)^2-(\iota')^2,  
\end{align}
where, at equilibrium, by \eqref{EL6}, $\Omega'=0$, considering the periodicity of the boundary, the quadratic term on the right-hand side of \eqref{var2omega2} can be expressed alternatively as
\begin{align}
\int_{\calC} (\del \Omega)^2 &=\int_{\calC} ((\kappa^{-1} \bfb \cdot \bfv'')'+\kappa \go \bfb \cdot \bfv'+\iota')^2 
\notag\\[4pt]
&= \int_{\calC}(((\kappa^{-1} \bfb \cdot \bfv'')'+\kappa \go \bfb \cdot \bfv')^2-(\iota')^2).
\label{var2omega3}
\end{align}
%%%%%%%%%%%%%%%%
%\section*{References}
\vspace{-4pt}
\bibliography{bib1} 
\bibliographystyle{elsarticle-harv}
\end{document}